\documentclass[aps,prd,reprint,preprintnumbers,showpacs,floatfix,nofootinbib,superscript address]{revtex4-2}

\usepackage[utf8]{inputenc}
\usepackage{parskip}
\usepackage{amssymb}
\usepackage{stix}
\usepackage{hhline}
\usepackage{amsmath}
\usepackage{mathtools}
\usepackage[dvipsnames]{xcolor}
\usepackage{xspace}
\usepackage{multirow,tabularx}
\usepackage{siunitx}
\usepackage{multirow}
\usepackage{graphicx}
\usepackage{xstring}
\usepackage{etoolbox}
\usepackage{notoccite}
\usepackage{tocbasic}
\usepackage{natbib}
\usepackage{mathrsfs}
\usepackage{lineno}
\usepackage{tensor}
\usepackage{accents}
\usepackage{tikz}
\usepackage{listings}
\allowdisplaybreaks

\usepackage{scalerel,stackengine}
\usepackage{upgreek}

\usepackage{enumitem}
\newrobustcmd{\Planck}{%
	{M_{\text{Pl}}}%
}

\newrobustcmd{\rD}[1]{%
	\tensor{\mathring{\nabla}}{#1}%
}

\newrobustcmd{\rcD}[1]{%
	\tensor{\nabla}{#1}%
}

\newrobustcmd{\rR}[1]{%
	\tensor{\mathring{R}}{#1}%
}

\newrobustcmd{\rcR}[1]{%
	\tensor{R}{#1}%
}

\newrobustcmd{\T}[2][placeholder]{%
	\IfEqCase{#1}{%
	{placeholder}{\tensor{T}{#2}}%
	{1}{\tensor[^{(1)}]{T}{#2}}%
	{2}{\tensor[^{(2)}]{T}{#2}}%
	{3}{\tensor[^{(3)}]{T}{#2}}%
	}%
	[\packageError{cosmicclass}{Symbol #1 is not an irreducible part!}{}]%
}

\newrobustcmd{\TLambda}[2][placeholder]{%
	\IfEqCase{#1}{%
	{placeholder}{\tensor{\lambda}{#2}}%
	{1}{\tensor[^{(1)}]{\lambda}{#2}}%
	{2}{\tensor[^{(2)}]{\lambda}{#2}}%
	{3}{\tensor[^{(3)}]{\lambda}{#2}}%
	}%
	[\packageError{cosmicclass}{Symbol #1 is not an irreducible part!}{}]%
}

\newrobustcmd{\TimeFunc}{	\mathscr{T}\xspace
}
\newrobustcmd{\SpaceFunc}{	\mathscr{R}\xspace
}

\newrobustcmd{\States}[4]{%
	{\tensor*{#4}{^#3_{#1^#2}}}%
}
\newrobustcmd{\FieldIndices}[2]{%
	{#2_#1}%
}
\newrobustcmd{\ParallelFieldIndices}[3]{%
	{{\overline{#3}}_{#1^#2}}%
}

\newrobustcmd{\FieldUp}[2]{%
	{\tensor{\zeta}{^{\FieldIndices{#1}{#2}}}}%
}
\newrobustcmd{\FieldDown}[2]{%
	{\tensor{\zeta}{_{\FieldIndices{#1}{#2}}}}%
}
\newrobustcmd{\SourceUp}[2]{%
	{\tensor{j}{^{\FieldIndices{#1}{#2}}}}%
}
\newrobustcmd{\SourceDown}[2]{%
	{\tensor{j}{_{\FieldIndices{#1}{#2}}}}%
}

\newrobustcmd{\FieldUpState}[5]{%
	{\tensor{\zeta\big(\States{#1}{#2}{#3}{#4}\big)}{^{\ParallelFieldIndices{#1}{#2}{#5}}}}%
}
\newrobustcmd{\SourceUpFullState}[5]{%
	{\tensor{j\big(\States{#1}{#2}{#3}{#4}\big)}{^{\FieldIndices{#3}{#5}}}}%
}
\newrobustcmd{\FieldDownState}[5]{%
	{\tensor{\zeta\big(\States{#1}{#2}{#3}{#4}\big)}{_{\ParallelFieldIndices{#1}{#2}{#5}}}}%
}
\newrobustcmd{\FieldDownFullState}[5]{%
	{\tensor{\zeta\big(\States{#1}{#2}{#3}{#4}\big)}{_{\FieldIndices{#3}{#5}}}}%
}

\newrobustcmd{\WaveOperatorTensorUpDown}[4]{%
	{\tensor{\mathcal{O}}{^{\FieldIndices{#1}{#3}}_{\FieldIndices{#2}{#4}}}}%
}

\newrobustcmd{\Normalisation}[4]{%
	{c\big(\States{#1}{#2}{#3}{#4}\big)}%
}
\newrobustcmd{\SPODownUp}[8]{%
	{\tensor{%
	\mathcal{P}%
	\big(\States{#1}{#2}{#3}{#5},\States{#1}{#2}{#4}{#6}\big)%
	}{_{\FieldIndices{#3}{#7}}^{\FieldIndices{#4}{#8}}}}%
}
\newrobustcmd{\SPOUpDown}[8]{%
	{\tensor{%
	\mathcal{P}%
	\big(\States{#1}{#2}{#3}{#5},\States{#1}{#2}{#4}{#6}\big)%
	}{^{\FieldIndices{#3}{#7}}_{\FieldIndices{#4}{#8}}}}%
}
\newrobustcmd{\ReducedSPODownUp}[6]{%
	{\tensor{%
	\mathcal{P}%
	\big(\States{#1}{#2}{#3}{#4}\big)%
	}{_{\ParallelFieldIndices{#1}{#2}{#5}}^{\FieldIndices{#3}{#6}}}}%
}
\newrobustcmd{\ReducedSPOUpDown}[6]{%
	{\tensor{%
	\mathcal{P}%
	\big(\States{#1}{#2}{#3}{#4}\big)%
	}{^{\ParallelFieldIndices{#1}{#2}{#5}}_{\FieldIndices{#3}{#6}}}}%
}

\newrobustcmd{\GaugeVarying}[3]{%
	{\tensor*{g}{_{#3_{#1^#2}}}}%
}
\newrobustcmd{\GaugeVaryingConj}[3]{%
	{\tensor*{g}{^*_{#3_{#1^#2}}}}%
}
\newrobustcmd{\GaugeFixing}[3]{%
	{\tensor*{h}{_{#3_{#1^#2}}}}%
}
\newrobustcmd{\Mass}[3]{%
	{\tensor*{m}{_{#3_{#1^#2}}}}%
}
\newrobustcmd{\SquareMass}[3]{%
	{\tensor*{m}{^2_{#3_{#1^#2}}}}%
}
\newrobustcmd{\NullVectors}[3]{%
	{#3_{#1^#2}}%
}
\newrobustcmd{\NullVector}[3]{%
	{\tensor*{\mathsf{v}}{_{#3_{#1^#2}}}}%
}

\newrobustcmd{\NullVectorComponents}[5]{%
	{\tensor{\left[\NullVector{#1}{#2}{#3}\right]}{_{\States{#1}{#2}{#4}{#5}}}}%
}

\newrobustcmd{\SpinMin}[0]{%
	{J_{\mathrm{F}}}%
}
\newrobustcmd{\SpinMax}[0]{%
	{J_{\mathrm{L}}}%
}

\newrobustcmd{\ParityMin}[0]{%
	{P_{\mathrm{F}}}%
}
\newrobustcmd{\ParityMax}[0]{%
	{P_{\mathrm{L}}}%
}

\newrobustcmd{\FieldMin}[0]{%
	{X_{\mathrm{F}}}%
}
\newrobustcmd{\FieldMax}[0]{%
	{X_{\mathrm{L}}}%
}

\newrobustcmd{\VectorMin}[0]{%
	{a_{\mathrm{F}}}%
}
\newrobustcmd{\VectorMax}[0]{%
	{a_{\mathrm{L}}}%
}

\newrobustcmd{\ZeroVector}[0]{%
	{\mathsf{0}}%
}
\newrobustcmd{\ConstraintMatrix}[0]{%
	{\mathsf{C}}%
}
\newrobustcmd{\SourceVector}[0]{%
	{\mathsf{J}}%
}

\newrobustcmd{\NullVectorConj}[3]{%
	{\tensor*{\mathsf{v}}{^\dagger_{#3_{#1^#2}}}}%
}

\newrobustcmd{\WaveOperator}[0]{%
	{\tensor{\mathsf{O}}{}}%
}
\newrobustcmd{\WaveOperatorConj}[0]{%
	{\tensor{\mathsf{O}}{^\dagger}}%
}
\newrobustcmd{\MoorePenrose}[0]{%
	{\tensor{\mathsf{O}}{^+}}%
}
\newrobustcmd{\MoorePenroseJP}[2]{%
	{\tensor*{\mathsf{O}}{^+_{#1^#2}}}%
}
\newrobustcmd{\WaveOperatorJP}[2]{%
	{\tensor*{\mathsf{O}}{_{#1^#2}}}%
}
\newrobustcmd{\WaveOperatorJPComponents}[6]{%
	{\tensor{\left[\WaveOperatorJP{#1}{#2}\right]}{_{\States{#1}{#2}{#3}{#5}\States{#1}{#2}{#4}{#6}}}}%
}
\newrobustcmd{\Propagator}[0]{%
	{\tensor{\mathsf{P}}{}}%
}
\newrobustcmd{\Similarity}[0]{%
	{\tensor{\mathsf{U}}{}}%
}

\newrobustcmd{\Source}[0]{%
	{\tensor{\mathsf{j}}{}}%
}
\newrobustcmd{\SourceConj}[0]{%
	{\tensor{\mathsf{j}}{^\dagger}}%
}
\newrobustcmd{\Field}[0]{%
	{\tensor{\upzeta}{}}%
}
\newrobustcmd{\FieldConj}[0]{%
	{\tensor{\upzeta}{^\dagger}}%
}

\newrobustcmd{\Powers}[3]{%
	{#3_{#1^#2}}%
}
\newrobustcmd{\Masses}[3]{%
	{#3_{#1^#2}}%
}

\newrobustcmd{\NewRes}[2]{%
	{\mathop{\mathrm{Res}}_{#1\mapsto#2}}%
}

\newrobustcmd{\DerOp}[2]{%
	\tensor[^{(#1)}]{\bar{\mathscr{D}}}{#2}%
}
\newrobustcmd{\bB}[1]{%
	\tensor{\bar{B}}{#1}%
}
\newrobustcmd{\bG}[1]{%
	\tensor*{\bar{G}}{#1}%
}
\newrobustcmd{\bg}[1]{%
	\tensor*{\bar{g}}{#1}%
}
\newrobustcmd{\bR}[1]{%
	\tensor*{\bar{R}}{#1}%
}
\newrobustcmd{\bD}[1]{%
	\tensor{\overline{\nabla}}{#1}%
}
\newrobustcmd{\hp}[1]{%
	\tensor*{h}{#1}%
}
\newrobustcmd{\hpb}[1]{%
	\tensor*{\bar{h}}{#1}%
}

\newrobustcmd{\Sources}[1]{%
	\hat{j}%
}
\newrobustcmd{\FieldsT}[1]{%
	\hat{\zeta}^{\text{T}}%
}

\newrobustcmd{\AMat}[1]{%
	{\mathsf{A}}%
}
\newrobustcmd{\AMatT}[1]{%
	{\mathsf{A}}^{\text{T}}%
}
\newrobustcmd{\AMatU}[1]{%
	\tensor[^{\text{s}}]{\mathsf{A}}{}%
}
\newrobustcmd{\AMatUt}[1]{%
	\tensor[^{\text{s}}]{\tilde{\mathsf{A}}}{}%
}
\newrobustcmd{\AMatO}[1]{%
	\tensor[^{\text{a}}]{\mathsf{A}}{}%
}

\newrobustcmd{\BMat}[1]{%
	{\mathsf{B}}%
}
\newrobustcmd{\BMatT}[1]{%
	{\mathsf{B}}^{\text{T}}%
}
\newrobustcmd{\BMatU}[1]{%
	\tensor[^{\text{s}}]{\mathsf{B}}{}%
}
\newrobustcmd{\BMatO}[1]{%
	\tensor[^{\text{a}}]{\mathsf{B}}{}%
}
\newrobustcmd{\BMatOt}[1]{%
	\tensor[^{\text{a}}]{\tilde{\mathsf{B}}}{}%
}

\newrobustcmd{\CMat}[1]{%
	{\mathsf{C}}%
}
\newrobustcmd{\CMatT}[1]{%
	{\mathsf{C}}^{\text{T}}%
}
\newrobustcmd{\CMatU}[1]{%
	\tensor[^{\text{s}}]{\mathsf{C}}{}%
}
\newrobustcmd{\CMatUt}[1]{%
	\tensor[^{\text{s}}]{\tilde{\mathsf{C}}}{}%
}

\newrobustcmd{\rotate}[1]{%
	\mathsf{R}%
}
\newrobustcmd{\rotateT}[1]{%
	\mathsf{R}^{\text{T}}%
}

\newrobustcmd{\omeg}[1]{%
	\mathsf{S}%
}
\newrobustcmd{\omegT}[1]{%
	\mathsf{S}^{\text{T}}%
}

\newrobustcmd{\uni}[1]{%
	\mathsf{n}_{#1}%
}
\newrobustcmd{\uniT}[1]{%
	\mathsf{n}_{#1}^{\text{T}}%
}
\newrobustcmd{\eig}[1]{%
	\mathsf{v}_{#1}%
}
\newrobustcmd{\eigT}[1]{%
	\mathsf{v}_{#1}^{\text{T}}%
}
\newrobustcmd{\lam}[1]{%
	\lambda_{#1}%
}

\newcommand{\mhDel}[1]{{\color[RGB]{0,100,0}\ifmmode\text{\sout{$#1$}}\else\sout{#1}\fi}}

\newrobustcmd{\pea}[1]{%
	\emph{#1}\textbf{.\ \ \ ---}
}

\newrobustcmd{\typeone}[1]{%
	{I}	
}
\newrobustcmd{\typetwo}[1]{%
	{II}	
}
\newrobustcmd{\typethree}[1]{%
	{III}	
}
\newrobustcmd{\typefour}[1]{%
	{IV}	
}
\newrobustcmd{\typefive}[1]{%
	{V}	
}
\newrobustcmd{\typesix}[1]{%
	{VI}	
}

\newrobustcmd{\first}[1]{%
	{1\textsuperscript{st}}	
}
\newrobustcmd{\second}[1]{%
	{2\textsuperscript{nd}}	
}
\newrobustcmd{\third}[1]{%
	{3\textsuperscript{rd}}	
}
\newrobustcmd{\fourth}[1]{%
	{4\textsuperscript{th}}	
}

\newrobustcmd{\IsOff}[1]{%
	{\ =0\ \land\ }	
}

\newrobustcmd{\MAGg}[1]{%
	\tensor{g}{#1}
}

\newrobustcmd{\MAGd}[1]{%
	\tensor*{\delta}{#1}
}
\newrobustcmd{\MAGl}[1]{%
	\tensor{\xi}{#1}
}

\newrobustcmd{\gflat}[1]{%
  \tensor{\eta}{#1}
}
\newrobustcmd{\h}[1]{%
  \tensor{h}{#1}
}
\newrobustcmd{\MAGA}[1]{%
  \tensor{A}{#1}
}
\newrobustcmd{\MAGF}[1]{%
	\tensor{\mathcal{F}}{#1}
}
\newrobustcmd{\MAGFP}[1]{%
	\tensor{\accentset{P}{\mathcal{F}}}{#1}
}
\newrobustcmd{\MAGFTri}[1]{%
	\tensor{\accentset{\Delta}{\mathcal{F}}}{#1}
}
\newrobustcmd{\MAGFa}[1]{%
	\tensor{\mathcal{F}}{^{(14)}#1}
}
\newrobustcmd{\MAGFb}[1]{%
	\tensor{\mathcal{F}}{^{(13)}#1}
}
\newrobustcmd{\MAGT}[1]{%
	\tensor{\mathcal{T}}{#1}
}
\newrobustcmd{\MAGQ}[1]{%
	\tensor{\mathcal{Q}}{#1}
}
\newrobustcmd{\MAGQt}[1]{%
	\tensor{\tilde{\mathcal{Q}}}{#1}
}

\newrobustcmd{\alp}[1]{%
       {\alpha}	
}

\newrobustcmd{\bet}[1]{%
  \tensor[^{(#1)}]{\mu}{}
}

\stackMath 
\newcommand\OmitIndices[1]{%
\savestack{\tmpbox}{\stretchto{%
\scaleto{%
\scalerel*[\widthof{\ensuremath{#1}}]{\kern-.6pt\curlywedge\kern-.6pt}%
{\rule[-\textheight/2]{1ex}{\textheight}}
}{\textheight}%
}{0.5ex}}%
\stackon[1pt]{#1}{\tmpbox}%
}

\newrobustcmd{\LPV}[1]{%
	\IfEqCase{#1}{%
	{}{L_{\text{PV}}}%
	}%
	[{L_{\text{PV}}\left(#1\right)}]%
}

\newrobustcmd{\F}[2][placeholder]{%
	\IfEqCase{#1}{%
	{placeholder}{\tensor{T}{#2}}%
	{1}{\tensor[^{(1)}]{F}{#2}}%
	{2}{\tensor[^{(2)}]{F}{#2}}%
	{3}{\tensor[^{(3)}]{F}{#2}}%
	}%
	[\packageError{cosmicclass}{Symbol #1 is not an irreducible part!}{}]%
}

\newrobustcmd{\GenericVector}[1]{%
	\smash{{#1}_{\tensor[^{{(2)}}]{\hspace{-1pt}\lambda}{}}^{J^P}}%
}
\newrobustcmd{\GenericTensor}[1]{%
	\smash{{#1}_{\tensor[^{{(1)}}]{\hspace{-1pt}\lambda}{}}^{J^P}}%
}

\newrobustcmd{\g}[1]{%
	\tensor{g}{#1}%
}
\newrobustcmd{\rcCon}[1]{%
	\tensor*{\Gamma}{#1}%
}
\newrobustcmd{\rCon}[1]{%
	\tensor*{\mathring{\Gamma}}{#1}%
}
\newrobustcmd{\B}[1]{%
	\tensor{B}{#1}%
}
\newrobustcmd{\PD}[1]{%
	\tensor{\partial}{#1}%
}
\newrobustcmd{\BConj}[1]{%
	\tensor{B^\dagger}{#1}%
}
\newrobustcmd{\N}[1]{%
	\tensor{n}{#1}%
}
\newrobustcmd{\J}[1]{%
	\tensor{J}{#1}%
}
\newrobustcmd{\En}[1]{%
	\mathcal{E}%
}
\newrobustcmd{\Mo}[1]{%
	p%
}
\newrobustcmd{\JConj}[1]{%
	\tensor{J^\dagger}{#1}%
}
\newrobustcmd{\K}[1]{%
	\tensor{X}{_#1}%
}
\newrobustcmd{\KConj}[1]{%
	\tensor*{X}{^{\dagger}_{#1}}%
}

\renewrobustcmd{\F}[2][placeholder]{%
	\IfEqCase{#1}{%
	{placeholder}{\tensor{T}{#2}}%
	{1}{\tensor[^{(1)}]{F}{#2}}%
	{2}{\tensor[^{(2)}]{F}{#2}}%
	{3}{\tensor[^{(3)}]{F}{#2}}%
	}%
	[\packageError{cosmicclass}{Symbol #1 is not an irreducible part!}{}]%
}

\newrobustcmd{\Si}[2][placeholder]{%
	\IfEqCase{#1}{%
	{placeholder}{\tensor{T}{#2}}%
	{1}{\tensor[^{(1)}]{S}{#2}}%
	{2}{\tensor[^{(2)}]{S}{#2}}%
	{3}{\tensor[^{(3)}]{S}{#2}}%
	}%
	[\packageError{cosmicclass}{Symbol #1 is not an irreducible part!}{}]%
}

\newrobustcmd{\mass}[2][placeholder]{%
	\IfEqCase{#1}{%
	{placeholder}{\tensor{m}{#2}}%
	{1}{\tensor[^{(1)}]{m}{#2}}%
	{2}{\tensor[^{(2)}]{m}{#2}}%
	{3}{\tensor[^{(3)}]{m}{#2}}%
	}%
	[\packageError{cosmicclass}{Symbol #1 is not an irreducible part!}{}]%
}

\newrobustcmd{\RLambda}[2][placeholder]{%
	\IfEqCase{#1}{%
	{placeholder}{\tensor{\lambda}{#2}}%
	{1}{\tensor[^1]{\lambda}{#2}}%
	{2}{\tensor[^2]{\lambda}{#2}}%
	{3}{\tensor[^3]{\lambda}{#2}}%
	{R}{\tensor[^{(R)}]{\lambda}{#2}}%
	}%
	[\packageError{cosmicclass}{Symbol #1 is not an irreducible part!}{}]%
}

\newrobustcmd{\QLambda}[2][placeholder]{%
	\IfEqCase{#1}{%
		{placeholder}{\tensor{\hat{\lambda}}{#2}}%
	{1}{\tensor[^1]{\hat{\lambda}}{#2}}%
	{2}{\tensor[^2]{\hat{\lambda}}{#2}}%
	{3}{\tensor[^3]{\hat{\lambda}}{#2}}%
	}%
	[\packageError{cosmicclass}{Symbol #1 is not an irreducible part!}{}]%
}

\newrobustcmd{\Mgra}[1]{%
  {\tensor{M}{_{\text{#1}}}}%
}
\newrobustcmd{\Mpro}[1]{%
  {\tensor{\mathproper{M}}{_{\text{#1}}}}%
}
\newrobustcmd{\Malt}[1]{%
  {\tensor{\mathscr{M}}{_{\text{#1}}}}%
}
\newrobustcmd{\Mkom}[1]{%
  {\tensor{\mathfrak{M}}{_{\text{#1}}}}%
}

\newrobustcmd{\Mtotal}{%
  {\tensor{M}{_{\text{T}}}}%
}

\newrobustcmd{\Qtotal}{%
  {\tensor{Q}{_{\text{T}}}}%
}

\newrobustcmd{\Qtotalcal}{%
  {\tensor{\mathcal{  Q}}{_{\text{T}}}}%
}

\newrobustcmd{\action}[1]{%
  {\tensor{S}{_{\text{#1}}}}%
}

\newrobustcmd{\lagrangian}[1]{%
  {\tensor{L}{_{\text{#1}}}}%
}

\newrobustcmd{\lagrangianprop}[1]{%
  {\tensor{\mathproper{L}}{_{\text{#1}}}}%
}

\newrobustcmd{\epl}{%
  {\tensor{\mathsf{e}}{_+}}%
}

\newrobustcmd{\epe}{%
  {\tensor{\mathsf{e}}{_\perp}}%
}

\newrobustcmd{\qz}{%
  {\text{\color{orange}\cmark}}%
}
\newrobustcmd{\jz}{%
  {\text{\color{red}\xmark}}%
}

\newrobustcmd{\projmatrix}[2][placeholder]{%
  {\tensor*{M}{_{#1}^{#2}}}
}
\newrobustcmd{\projorthhum}[2][placeholder]{%
  {\tensor[^#2]{\smash{\check{\mathcal{  P}}}}{#1}}
}
\newrobustcmd{\projorthhumu}[2][placeholder]{%
  {\tensor[^#2]{\smash{\check{\mathcal{  P}}}}{#1}}
}
\newrobustcmd{\projorth}[2][placeholder]{%
  {\tensor[^#2]{\smash{\hat{\mathcal{  P}}}}{#1}}
}
\newrobustcmd{\projlore}[2][placeholder]{%
  {\tensor[^#2]{\hat{\mathcal{  P}}}{#1}}
}

\newrobustcmd{\gensec}[3][placeholder]{%
  {\tensor*[^#1]{\chi}{^{#2}_{\acu{#3}}}}
}

\newrobustcmd{\glfourr}{%
  {\mathrm{GL}(4,\mathbb{R})}%
}

\newrobustcmd{\sltwoc}{%
  {\mathrm{SL}(2,\mathbb{C})}%
}

\newrobustcmd{\poincare}{%
  {\mathbb{R}^{1,3}\rtimes\mathrm{SO}^+(1,3)}%
}

\newrobustcmd{\poincaref}{%
  {\mathrm{P}(1,3)}%
}

\newrobustcmd{\weyl}{%
  {\mathrm{W}(1,3)}%
}

\newrobustcmd{\conformal}{%
  {\mathrm{C}(1,3)}%
}

\newrobustcmd{\diffeomorphism}{%
  {\mathbb{R}^{1,3}}%
}

\newrobustcmd{\soonethree}{%
  {\mathrm{SO}^+(1,3)}%
}

\newrobustcmd{\othree}{%
  {\mathrm{SO}(3)}%
}

\newrobustcmd{\sothree}{%
  {\mathrm{SO}(3)}%
}

\newrobustcmd{\sotwo}{%
  {\mathrm{SO}(2)}%
}

\newrobustcmd{\suthreec}{%
  {\mathrm{SU}(3)_{\text{c}}}%
}

\newrobustcmd{\sutwol}{%
  {\mathrm{SU}(2)_{\text{L}}}%
}

\newrobustcmd{\uoney}{%
  {\mathrm{U}(1)_{\text{Y}}}%
}

\newrobustcmd{\uone}{%
  {\mathrm{U}(1)}%
}

\newrobustcmd{\uoneem}{%
  {\mathrm{U}(1)_{\text{em}}}%
}

\newrobustcmd{\sutwo}{%
  {\mathrm{SU}(2)}%
}

\newrobustcmd{\eplus}{%
  {\tensor{\mathsf{e}}{_{+}}}%
}

\newrobustcmd{\esf}[1]{%
  {\tensor{\mathsf{e}}{_{#1}}}
}%
\newrobustcmd{\esfu}[1]{%
  {\tensor{\mathsf{e}}{^{#1}}}
}%
\newrobustcmd{\gam}[1]{%
  {\tensor{\gamma}{_{#1}}}
}%
\newrobustcmd{\gamu}[1]{%
  {\tensor{\gamma}{^{#1}}}
}%

\newrobustcmd{\planck}{%
  {m_{\text{p}}}%
}

\newrobustcmd{\Pg}{%
	{\Phi_{\text{Nt}}}%
}
\newrobustcmd{\Rh}{%
	{r_{\text{Th}}}%
}
\newrobustcmd{\onshell}{%
	{\ =\ }%
}
\newrobustcmd{\Hl}{%
	{h_{\gamma}}%
}
\newrobustcmd{\Kl}{%
	{k_{\gamma}}%
}
\newrobustcmd{\Rb}{%
	{\rho_{\text{Br}}}%
}
\newrobustcmd{\Amond}{%
	{a_{0}}%
}
\newrobustcmd{\Anew}{%
	{a_{\text{Nt}}}%
}
\newrobustcmd{\Aobs}{%
	{a_{\text{Ob}}}%
}
\newrobustcmd{\Rl}{%
	{r_{\gamma}}%
}
\newrobustcmd{\Rp}{%
	{r_{+}}%
}
\newrobustcmd{\Rm}{%
	{r_{-}}%
}
\newrobustcmd{\Risco}{%
	{r_{\pm}}%
}
\newrobustcmd{\Rg}{%
	{r_{\text{Sz}}}%
}
\newrobustcmd{\Kb}{%
	{K_{\text{B}}}%
}

\newrobustcmd{\caligR}{%
  {\mathcal{R}}%
}
\newrobustcmd{\caligT}{%
  {\mathcal{T}}%
}

\newrobustcmd{\pgt}{%
  PGT\textsuperscript{q,+}\ %
}

\newrobustcmd{\unl}[1]{%
  {\mathfrak{#1}}%
}
\newrobustcmd{\ovl}[1]{%
\overline{#1}%
}
\newrobustcmd{\acu}[1]{%
\acute{#1}%
}

\newrobustcmd{\indiq}[2][placeholder]{%
\IfEqCase{#1}{%
  {placeholder}{%
    \IfEqCase{#2}{%
      {1}{\ovl{k}}%
      {2}{\ovl{kl}}%
      {3}{\ovl{klm}}%
    }%
  }%
}[#1]%
}%

\newrobustcmd{\indaq}[2][placeholder]{%
\IfEqCase{#1}{%
  {placeholder}{%
    \IfEqCase{#2}{%
      {1}{\overline{k}}%
      {2}{\overline{kl}}%
      {3}{\overline{klm}}%
    }%
  }%
}[#1]%
}%

\newrobustcmd{\indeq}[2][placeholder]{%
\IfEqCase{#1}{%
  {placeholder}{%
    \IfEqCase{#2}{%
      {1}{k}%
      {2}{kl}%
      {3}{klm}%
    }%
  }%
}[#1]%
}%

\newrobustcmd{\indoq}[2][placeholder]{%
\IfEqCase{#1}{%
  {placeholder}{%
    \IfEqCase{#2}{%
      {1}{\alpha}%
      {2}{\alpha\beta}%
      {3}{\alpha\beta\gamma}%
    }%
  }%
}[#1]%
}%

\newrobustcmd{\fcphi}[1]{%
  \tensor[^{\text{FC}}]{\phi}{_{#1}}%
}

\newrobustcmd{\scphi}[1]{%
  \tensor[^{\text{SC}}]{\phi}{_{#1}}%
}

\newrobustcmd{\fcmul}[1]{%
  \tensor[^{\text{FC}}]{\upsilon}{_{#1}}%
}

\newrobustcmd{\arb}{%
  {\tensor{f}{_{\text{lin}}}}%
}

\newrobustcmd{\scmul}[1]{%
  \tensor[^{\text{SC}}]{\upsilon}{_{#1}}%
}

\newrobustcmd{\foli}[1]{%
\tensor{n}{_{#1}}%
}

\newrobustcmd{\foliu}[1]{%
\tensor{n}{^{#1}}%
}

\newrobustcmd{\covderl}[1]{%
\tensor{\mathcal{D}}{^{\flat}_{\indiq[#1]{1}}}%
}

\newrobustcmd{\covder}[1]{%
\tensor{\mathcal{D}}{_{\indiq[#1]{1}}}%
}
\newrobustcmd{\coder}[1]{%
\tensor{D}{_{\indiq[#1]{1}}}%
}

\newrobustcmd{\deltal}[2]{%
  \tensor*{\delta}{_{\phantom{\flat}}^{\flat}_{#1}^{#2}}%
}

\newrobustcmd{\deltaud}[2]{%
  \tensor*{\delta}{^{#1}_{#2}}%
}
\newrobustcmd{\etau}[1]{%
\tensor{\eta}{^{\indiq[#1]{2}}}%
}
\newrobustcmd{\etaul}[1]{%
\tensor{\eta}{^{\flat}^{\indiq[#1]{2}}}%
}
\newrobustcmd{\etad}[1]{%
\tensor{\eta}{_{\indiq[#1]{2}}}%
}
\newrobustcmd{\etadl}[1]{%
\tensor{\eta}{^{\flat}_{\indiq[#1]{2}}}%
}

\newrobustcmd{\epsul}[1]{%
\tensor{\epsilon}{^{\flat}^{\indiq[#1]{3}}^{\perp}}
}
\newrobustcmd{\epsdl}[1]{%
\tensor{\epsilon}{^{\flat}_{\indiq[#1]{3}}_{\perp}}
}
\newrobustcmd{\epsd}[1]{%
\tensor{\epsilon}{_{\indiq[#1]{3}}_{\perp}}
}
\newrobustcmd{\epsu}[1]{%
\tensor{\epsilon}{^{\indiq[#1]{3}}^{\perp}}
}

\newrobustcmd{\hfl}[2]{%
  \tensor{h}{^{\flat}_{#1}^{#2}}
}

\newrobustcmd{\cgalp}{\tensor{\alpha}{_{\text{CG}}}}

\newrobustcmd{\cbet}[1]{%
  \tensor{\bar{\beta}}{_{#1}}
}
\newrobustcmd{\calp}[1]{%
  \tensor{\bar{\alpha}}{_{#1}}
}

\newrobustcmd{\alpg}[1]{%
  \tensor{\check{\alpha}}{_{#1}}
}
\newrobustcmd{\betg}[1]{%
  \tensor{\check{\beta}}{_{#1}}
}
\newrobustcmd{\calpg}[1]{%
  \tensor{\acu{\alpha}}{_{#1}}
}
\newrobustcmd{\cbetg}[1]{%
  \tensor{\acu{\beta}}{_{#1}}
}

\newrobustcmd{\hub}{%
  {\underline{\mathsf{h}}}
}
\newrobustcmd{\hubm}{%
  {\underline{\mathsf{h}}^{-1}}
}
\newrobustcmd{\hob}{%
  {\bar{\mathsf{h}}}
}
\newrobustcmd{\hobm}{%
  {\bar{\mathsf{h}}^{-1}}
}
\newrobustcmd{\hdet}{%
  {\det \mathsf{h}}
}
\newrobustcmd{\hmdet}{%
  {\det \mathsf{h}^{-1}}
}
\newrobustcmd{\Rsf}{%
  {\mathsf{R}}
}

\newrobustcmd{\alpm}[2][placeholder]{%
  {\tensor*{\hat{\alpha}}{_{#1}^{#2}}}
}
\newrobustcmd{\calpm}[2][placeholder]{%
  \tensor*{\bar{\alpha}}{_{#1}^{#2}}
}

\newrobustcmd{\betm}[2][placeholder]{%
  {\tensor*{\hat{\beta}}{_{#1}^{#2}}}
}
\newrobustcmd{\cbetm}[2][placeholder]{%
  \tensor*{\bar{\beta}}{_{#1}^{#2}}
}

\newrobustcmd{\lamr}{%
  {\zeta_{\mathcal{  R}} }
}
\newrobustcmd{\barlamr}{%
  {\bar{\zeta}_{\mathcal{  R}} }
}
\newrobustcmd{\lamt}{%
  {\zeta_{\mathcal{  T}} }
}
\newrobustcmd{\barlamt}{%
  {\bar{\zeta}_{\mathcal{  T}} }
}

\newrobustcmd{\atmp}[1]{%
  \tensor{\hat{a}}{_{#1}}
}
\newrobustcmd{\btmp}[1]{%
  \tensor{b}{_{#1}}
}

\newrobustcmd{\ctmp}[2][placeholder]{%
  {\tensor*{c}{_{#1}^{#2}}}
}
\newrobustcmd{\dtmp}[2][placeholder]{%
  {\tensor*{d}{_{#1}^{#2}}}
}

\newrobustcmd{\etmp}[1]{%
  \tensor{e}{_{#1}}
}

\newrobustcmd{\batmp}[1]{%
  \tensor{\ovl{a}}{_{#1}}
}
\newrobustcmd{\bbtmp}[1]{%
  \tensor{\ovl{b}}{_{#1}}
}
\newrobustcmd{\bctmp}[1]{%
  \tensor{\ovl{c}}{_{#1}}
}
\newrobustcmd{\bdtmp}[1]{%
  \tensor{\ovl{d}}{_{#1}}
}
\newrobustcmd{\betmp}[1]{%
  \tensor{\ovl{e}}{_{#1}}
}

\newrobustcmd{\ptl}[1]{%
  \tensor{\partial}{#1}
}

\newrobustcmd{\etaf}[1]{%
  \tensor{\eta}{#1}
}

\newrobustcmd{\epsf}[1]{%
  \tensor{\epsilon}{#1}
}

\newrobustcmd{\RSO}[2][placeholder]{%
  {\tensor[^{#2}]{\mathcal{  R}}{#1}}
}
\newrobustcmd{\TSO}[2][placeholder]{%
  {\tensor[^{#2}]{\mathcal{  T}}{#1}}
}
\newrobustcmd{\FSO}[2][placeholder]{%
  {\tensor[^{#2}]{\mathcal{  F}}{#1}}
}
\newrobustcmd{\spinSO}[2][placeholder]{%
  {\tensor[^{#2}]{\sigma}{#1}}
}
\newrobustcmd{\RLambdaSO}[2][placeholder]{%
  {\tensor[^{#2}]{\zeta}{#1}}
}
\newrobustcmd{\TLambdaSO}[2][placeholder]{%
  {\tensor[^{#2}]{\zeta}{#1}}
}
\newrobustcmd{\KSO}[2][placeholder]{%
  {\tensor[^{#2}]{\mathcal{  K}}{#1}}
}

\newrobustcmd{\bper}[2][placeholder]{%
\IfEqCase{#2}{%
  {s}{\tensor{\mathfrak{s}}{#1}}%
  {a}{\tensor{\mathfrak{a}}{#1}}%
  {sbar}{\tensor{\bar{\mathfrak{s}}}{#1}}%
}[\packageError{cosmicclass}{Unidentified Critical Case: #1}{}]%
}

\newrobustcmd{\Jl}{%
  {J^{\flat}}%
}%

\newrobustcmd{\Nl}{%
  {N^{\flat}}%
}%

\newrobustcmd{\haml}[2][placeholder]{%
\IfEqCase{#2}{%
{mom0p}{\tensor{\mathcal{H}}{^{\flat}_{\perp}}}%
{mom1m}{\tensor{\mathcal{H}}{^{\flat}_{\indoq[#1]{1}}}}%
{rot1p}{\tensor{\mathcal{H}}{^{\flat}_{\indaq[#1]{2}}}}%
{rot1m}{\tensor{\mathcal{H}}{^{\flat}_{\perp}_{\indaq[#1]{1}}}}%
}[\packageError{cosmicclass}{Unidentified Critical Case: #1}{}]%
}

\newrobustcmd{\arc}[2][placeholder]{%
\IfEqCase{#2}{%
{B1p}{\tensor{\vartheta}{_{\perp\indiq[#1]{2}}}}%
{B2m}{\tensor[^{\text{T}}]{\vartheta}{_{\indiq[#1]{3}}}}%
{A0m}{\tensor[^{\text{P}}]{\vartheta}{}}%
{A1p}{\tensor{\overset{\wedge}{\vartheta}}{_{\perp\indiq[#1]{2}}}}%
{A1m}{\tensor{\overset{\rightharpoonup}{\vartheta}}{_{\indiq[#1]{1}}}}%
{A2p}{\tensor{\overset{\sim}{\vartheta}}{_{\perp\indiq[#1]{2}}}}%
{A2m}{\tensor[^{\text{T}}]{\vartheta}{_{\perp\indiq[#1]{3}}}}%
}[\packageError{cosmicclass}{Unidentified Critical Case: #1}{}]%
}

\newrobustcmd{\pic}[2][placeholder]{%
\IfEqCase{#2}{%
{B0p}{\varphi}%
{B1p}{\tensor{\overset{\wedge}{\varphi}}{_{\indiq[#1]{2}}}}%
{B1m}{\tensor{\varphi}{_{\perp\indiq[#1]{1}}}}%
{B2p}{\tensor{\overset{\sim}{\varphi}}{_{\indiq[#1]{2}}}}%
{A0p}{\tensor{\varphi}{_\perp}}%
{A0m}{\tensor[^{\text{P}}]{\varphi}{}}%
{A1p}{\tensor{\overset{\wedge}{\varphi}}{_{\perp\indiq[#1]{2}}}}%
{A1m}{\tensor{\overset{\rightharpoonup}{\varphi}}{_{\indiq[#1]{1}}}}%
{A2p}{\tensor{\overset{\sim}{\varphi}}{_{\perp\indiq[#1]{2}}}}%
{A2m}{\tensor[^{\text{T}}]{\varphi}{_{\indiq[#1]{3}}}}%
}[\packageError{cosmicclass}{Unidentified Critical Case: #1}{}]%
}

\newrobustcmd{\picu}[2][placeholder]{%
\IfEqCase{#2}{%
{B0p}{\varphi}%
{B1p}{\tensor{\smash{\overset{\wedge}{\varphi}}}{^{\indiq[#1]{2}}}}%
{B1m}{\tensor{\varphi}{^{\perp\indiq[#1]{1}}}}%
{B2p}{\tensor{\smash{\overset{\sim}{\varphi}}}{^{\indiq[#1]{2}}}}%
{A0p}{\tensor{\varphi}{_\perp}}%
{A0m}{\tensor[^{\text{P}}]{\varphi}{}}%
{A1p}{\tensor{\smash{\overset{\wedge}{\varphi}}}{^{\perp\indiq[#1]{2}}}}%
{A1m}{\tensor{\smash{\overset{\rightharpoonup}{\varphi}}}{^{\indiq[#1]{1}}}}%
{A2p}{\tensor{\smash{\overset{\sim}{\varphi}}}{^{\perp\indiq[#1]{2}}}}%
{A2m}{\tensor[^{\text{T}}]{\varphi}{^{\indiq[#1]{3}}}}%
}[\packageError{cosmicclass}{Unidentified Critical Case: #1}{}]%
}

\newrobustcmd{\picl}[2][placeholder]{%
\IfEqCase{#2}{%
{B0p}{\tensor{\varphi}{^{\flat}}}%
{B1p}{\tensor{\smash{\overset{\wedge}{\varphi}}}{^{\flat}_{\indiq[#1]{2}}}}%
{B1m}{\tensor{\varphi}{^{\flat}_{\perp}_{\indiq[#1]{1}}}}%
{B2p}{\tensor{\smash{\overset{\sim}{\varphi}}}{^{\flat}_{\indiq[#1]{2}}}}%
{A0p}{\tensor{\varphi}{_\perp}^{\flat}}%
{A0m}{\tensor[^{\text{P}}]{\varphi}{^{\flat}}}%
{A1p}{\tensor{\smash{\overset{\wedge}{\varphi}}}{^{\flat}_{\perp\indiq[#1]{2}}}}%
{A1m}{\tensor{\smash{\overset{\rightharpoonup}{\varphi}}}{^{\flat}_{\indiq[#1]{1}}}}%
{A2p}{\tensor{\smash{\overset{\sim}{\varphi}}}{^{\flat}_{\perp\indiq[#1]{2}}}}%
{A2m}{\tensor[^{\text{T}}]{\varphi}{^{\flat}_{\indiq[#1]{3}}}}%
}[\packageError{cosmicclass}{Unidentified Critical Case: #1}{}]%
}

\newrobustcmd{\mull}[2][placeholder]{%
\IfEqCase{#2}{%
{B0p}{\tensor{u}{^{\flat}}}%
{B1p}{\tensor{\smash{\overset{\wedge}{u}}}{^{\flat}_{\indiq[#1]{2}}}}%
{B1m}{\tensor{u}{^{\flat}_{\perp}_{\indiq[#1]{1}}}}%
{B2p}{\tensor{\smash{\overset{\sim}{u}}}{^{\flat}_{\indiq[#1]{2}}}}%
{A0p}{\tensor{u}{_\perp}^{\flat}}%
{A0m}{\tensor[^{\text{P}}]{u}{^{\flat}}}%
{A1p}{\tensor{\smash{\overset{\wedge}{u}}}{^{\flat}_{\perp\indiq[#1]{2}}}}%
{A1m}{\tensor{\smash{\overset{\rightharpoonup}{u}}}{^{\flat}_{\indiq[#1]{1}}}}%
{A2p}{\tensor{\smash{\overset{\sim}{u}}}{^{\flat}_{\perp\indiq[#1]{2}}}}%
{A2m}{\tensor[^{\text{T}}]{u}{^{\flat}_{\indiq[#1]{3}}}}%
}[\packageError{cosmicclass}{Unidentified Critical Case: #1}{}]%
}

\newrobustcmd{\mul}[2][placeholder]{%
\IfEqCase{#2}{%
{B0p}{\tensor{u}{}}%
{B1p}{\tensor{\smash{\overset{\wedge}{u}}}{_{\indiq[#1]{2}}}}%
{B1m}{\tensor{u}{_{\perp}_{\indiq[#1]{1}}}}%
{B2p}{\tensor{\smash{\overset{\sim}{u}}}{_{\indiq[#1]{2}}}}%
{A0p}{\tensor{u}{_\perp}}%
{A0m}{\tensor[^{\text{P}}]{u}{}}%
{A1p}{\tensor{\smash{\overset{\wedge}{u}}}{_{\perp\indiq[#1]{2}}}}%
{A1m}{\tensor{\smash{\overset{\rightharpoonup}{u}}}{_{\indiq[#1]{1}}}}%
{A2p}{\tensor{\smash{\overset{\sim}{u}}}{_{\perp\indiq[#1]{2}}}}%
{A2m}{\tensor[^{\text{T}}]{u}{_{\indiq[#1]{3}}}}%
}[\packageError{cosmicclass}{Unidentified Critical Case: #1}{}]%
}

\newrobustcmd{\PiP}[2][placeholder]{%
\IfEqCase{#2}{%
{B0p}{\hat{\pi}}%
{B1p}{\tensor{\overset{\wedge}{\hat{\pi}}}{_{\indiq[#1]{2}}}}%
{B1m}{\tensor{\hat{\pi}}{_{\perp\indiq[#1]{1}}}}%
{B2p}{\tensor{\overset{\sim}{\hat{\pi}}}{_{\indiq[#1]{2}}}}%
{A0p}{\tensor{\hat{\pi}}{_\perp}}%
{A0m}{\tensor[^{\text{P}}]{\hat{\pi}}{}}%
{A1p}{\tensor{\overset{\wedge}{\hat{\pi}}}{_{\perp\indiq[#1]{2}}}}%
{A1m}{\tensor{\overset{\rightharpoonup}{\hat{\pi}}}{_{\indiq[#1]{1}}}}%
{A2p}{\tensor{\overset{\sim}{\hat{\pi}}}{_{\perp\indiq[#1]{2}}}}%
{A2m}{\tensor[^{\text{T}}]{\hat{\pi}}{_{\indiq[#1]{3}}}}%
}[\packageError{cosmicclass}{Unidentified Critical Case: #1}{}]%
}

\newrobustcmd{\PiPu}[2][placeholder]{%
\IfEqCase{#2}{%
{B0p}{\hat{\pi}}%
{B1p}{\tensor{\smash{\overset{\wedge}{\hat{\pi}}}}{^{\indiq[#1]{2}}}}%
{B1m}{\tensor{\smash{\hat{\pi}}}{^{\perp\indiq[#1]{1}}}}%
{B2p}{\tensor{\smash{\overset{\sim}{\hat{\pi}}}}{^{\indiq[#1]{2}}}}%
{A0p}{\tensor{\smash{\hat{\pi}}}{^\perp}}%
{A0m}{\tensor[^{\text{P}}]{\smash{\hat{\pi}}}{}}%
{A1p}{\tensor{\smash{\overset{\wedge}{\hat{\pi}}}}{^{\perp\indiq[#1]{2}}}}%
{A1m}{\tensor{\smash{\overset{\rightharpoonup}{\hat{\pi}}}}{^{\indiq[#1]{1}}}}%
{A2p}{\tensor{\smash{\overset{\sim}{\hat{\pi}}}}{^{\perp\indiq[#1]{2}}}}%
{A2m}{\tensor[^{\text{T}}]{\smash{\hat{\pi}}}{^{\indiq[#1]{3}}}}%
}[\packageError{cosmicclass}{Unidentified Critical Case: #1}{}]%
}

\newrobustcmd{\sicl}[2][placeholder]{%
\IfEqCase{#2}{%
{B0p}{\tensor{\chi}{^{\flat}}}%
{B1p}{\tensor{\smash{\overset{\wedge}{\chi}}}{^{\flat}_{\indiq[#1]{2}}}}%
{B1m}{\tensor{\chi}{^{\flat}_{\perp}_{\indiq[#1]{1}}}}%
{B2p}{\tensor{\smash{\overset{\sim}{\chi}}}{^{\flat}_{\indiq[#1]{2}}}}%
{A0p}{\tensor{\chi}{^{\flat}_\perp}}%
{A0m}{\tensor[^{\text{P}}]{\chi}{^{\flat}}}%
{A1p}{\tensor{\smash{\overset{\wedge}{\chi}}}{^{\flat}_{\perp\indiq[#1]{2}}}}%
{A1m}{\tensor{\smash{\overset{\rightharpoonup}{\chi}}}{^{\flat}_{\indiq[#1]{1}}}}%
{A2p}{\tensor{\smash{\overset{\sim}{\chi}}}{^{\flat}_{\perp\indiq[#1]{2}}}}%
{A2m}{\tensor[^{\text{T}}]{\chi}{^{\flat}_{\indiq[#1]{3}}}}%
}[\packageError{cosmicclass}{Unidentified Critical Case: #1}{}]%
}

\newrobustcmd{\ticl}[2][placeholder]{%
\IfEqCase{#2}{%
{B0p}{\tensor{\zeta}{^{\flat}}}%
{B1p}{\tensor{\smash{\overset{\wedge}{\zeta}}}{^{\flat}_{\indiq[#1]{2}}}}%
{B1m}{\tensor{\zeta}{^{\flat}_{\perp}_{\indiq[#1]{1}}}}%
{B2p}{\tensor{\smash{\overset{\sim}{\zeta}}}{^{\flat}_{\indiq[#1]{2}}}}%
{A0p}{\tensor{\zeta}{^{\flat}_\perp}}%
{A0m}{\tensor[^{\text{P}}]{\zeta}{^{\flat}}}%
{A1p}{\tensor{\smash{\overset{\wedge}{\zeta}}}{^{\flat}_{\perp\indiq[#1]{2}}}}%
{A1m}{\tensor{\smash{\overset{\rightharpoonup}{\zeta}}}{^{\flat}_{\indiq[#1]{1}}}}%
{A2p}{\tensor{\smash{\overset{\sim}{\zeta}}}{^{\flat}_{\perp\indiq[#1]{2}}}}%
{A2m}{\tensor[^{\text{T}}]{\zeta}{^{\flat}_{\indiq[#1]{3}}}}%
}[\packageError{cosmicclass}{Unidentified Critical Case: #1}{}]%
}

\newrobustcmd{\PiPl}[2][placeholder]{%
\IfEqCase{#2}{%
{B0p}{\tensor{\hat{\pi}}{^{\flat}}}%
{B1p}{\tensor{\smash{\overset{\wedge}{\hat{\pi}}}}{^{\flat}_{\indiq[#1]{2}}}}%
{B1m}{\tensor{\hat{\pi}}{^{\flat}_{\perp}_{\indiq[#1]{1}}}}%
{B2p}{\tensor{\smash{\overset{\sim}{\hat{\pi}}}}{^{\flat}_{\indiq[#1]{2}}}}%
{A0p}{\tensor{\hat{\pi}}{_\perp}^{\flat}}%
{A0m}{\tensor[^{\text{P}}]{\hat{\pi}}{^{\flat}}}%
{A1p}{\tensor{\smash{\overset{\wedge}{\hat{\pi}}}}{^{\flat}_{\perp\indiq[#1]{2}}}}%
{A1m}{\tensor{\smash{\overset{\rightharpoonup}{\hat{\pi}}}}{^{\flat}_{\indiq[#1]{1}}}}%
{A2p}{\tensor{\smash{\overset{\sim}{\hat{\pi}}}}{^{\flat}_{\perp\indiq[#1]{2}}}}%
{A2m}{\tensor[^{\text{T}}]{\hat{\pi}}{^{\flat}_{\indiq[#1]{3}}}}%
}[\packageError{cosmicclass}{Unidentified Critical Case: #1}{}]%
}

\newrobustcmd{\sic}[2][placeholder]{%
\IfEqCase{#2}{%
{B0p}{\chi}%
{B1p}{\tensor{\overset{\wedge}{\chi}}{_{\indiq[#1]{2}}}}%
{B1m}{\tensor{\chi}{_{\perp\indiq[#1]{1}}}}%
{B2p}{\tensor{\overset{\sim}{\chi}}{_{\indiq[#1]{2}}}}%
{A0p}{\tensor{\chi}{_\perp}}%
{A0m}{\tensor[^{\text{P}}]{\chi}{}}%
{A1p}{\tensor{\overset{\wedge}{\chi}}{_{\perp\indiq[#1]{2}}}}%
{A1m}{\tensor{\overset{\rightharpoonup}{\chi}}{_{\indiq[#1]{1}}}}%
{A2p}{\tensor{\overset{\sim}{\chi}}{_{\perp\indiq[#1]{2}}}}%
{A2m}{\tensor[^{\text{T}}]{\chi}{_{\indiq[#1]{3}}}}%
}[\packageError{cosmicclass}{Unidentified Critical Case: #1}{}]%
}

\newrobustcmd{\lorsicpar}[2][placeholder]{%
\IfEqCase{#2}{%
{B0m}{\tensor*[^{\text{P}}]{\smash{\underline{\chi}}}{^{\parallel}}}%
{B1p}{\tensor*{\smash{\overset{\wedge}{\chi}}}{^{\parallel}_{\indiq[#1]{2}}}}%
{B1m}{\tensor*{\chi}{^{\parallel}_{\perp\indiq[#1]{1}}}}%
{B2m}{\tensor*[^{\text{T}}]{\smash{\underline{\chi}}}{^{\parallel}_{\indiq[#1]{3}}}}%
{A0p}{\tensor*{\chi}{^{\parallel}_\perp}}%
{A0m}{\tensor*[^{\text{P}}]{\chi}{^{\parallel}}}%
{A1p}{\tensor*{\smash{\overset{\wedge}{\chi}}}{^{\parallel}_{\perp\indiq[#1]{2}}}}%
{A1m}{\tensor*{\smash{\overset{\rightharpoonup}{\chi}}}{^{\parallel}_{\indiq[#1]{1}}}}%
{A2p}{\tensor*{\smash{\overset{\sim}{\chi}}}{^{\parallel}_{\perp\indiq[#1]{2}}}}%
{A2m}{\tensor*[^{\text{T}}]{\chi}{^{\parallel}_{\indiq[#1]{3}}}}%
}[\packageError{cosmicclass}{Unidentified Critical Case: #1}{}]%
}

\newrobustcmd{\lorsicpir}[2][placeholder]{%
\IfEqCase{#2}{%
{B0p}{\tensor*{\chi}{^{\vDash}}}%
{B1p}{\tensor*{\smash{\overset{\wedge}{\chi}}}{^{\vDash}_{\indiq[#1]{2}}}}%
{B1m}{\tensor*{\chi}{^{\vDash}_{\perp\indiq[#1]{1}}}}%
{B2p}{\tensor*{\smash{\overset{\sim}{\chi}}}{^{\vDash}_{\indiq[#1]{2}}}}%
{A0p}{\tensor*{\chi}{^{\vDash}_\perp}}%
{A0m}{\tensor*[^{\text{P}}]{\chi}{^{\vDash}}}%
{A1p}{\tensor*{\smash{\overset{\wedge}{\chi}}}{^{\vDash}_{\perp\indiq[#1]{2}}}}%
{A1m}{\tensor*{\smash{\overset{\rightharpoonup}{\chi}}}{^{\vDash}_{\indiq[#1]{1}}}}%
{A2p}{\tensor*{\smash{\overset{\sim}{\chi}}}{^{\vDash}_{\perp\indiq[#1]{2}}}}%
{A2m}{\tensor*[^{\text{T}}]{\chi}{^{\vDash}_{\indiq[#1]{3}}}}%
}[\packageError{cosmicclass}{Unidentified Critical Case: #1}{}]%
}

\newrobustcmd{\lorsicper}[2][placeholder]{%
\IfEqCase{#2}{%
{B0p}{\tensor*{\chi}{^{\perp}}}%
{B1p}{\tensor*{\smash{\overset{\wedge}{\chi}}}{^{\perp}_{\indiq[#1]{2}}}}%
{B1m}{\tensor*{\chi}{^{\perp}_{\perp\indiq[#1]{1}}}}%
{B2p}{\tensor*{\smash{\overset{\sim}{\chi}}}{^{\perp}_{\indiq[#1]{2}}}}%
{A0p}{\tensor*{\chi}{^{\perp}_\perp}}%
{A0m}{\tensor*[^{\text{P}}]{\chi}{^{\perp}}}%
{A1p}{\tensor*{\smash{\overset{\wedge}{\chi}}}{^{\perp}_{\perp\indiq[#1]{2}}}}%
{A1m}{\tensor*{\smash{\overset{\rightharpoonup}{\chi}}}{^{\perp}_{\indiq[#1]{1}}}}%
{A2p}{\tensor*{\smash{\overset{\sim}{\chi}}}{^{\perp}_{\perp\indiq[#1]{2}}}}%
{A2m}{\tensor*[^{\text{T}}]{\chi}{^{\perp}_{\indiq[#1]{3}}}}%
}[\packageError{cosmicclass}{Unidentified Critical Case: #1}{}]%
}

\newrobustcmd{\Tl}[2][placeholder]{%
\IfEqCase{#2}{%
{B0p}{\tensor{\chi}{^{\flat}}}%
{B1p}{\tensor{\smash{\overset{\wedge}{\chi}}}{^{\flat}_{\indiq[#1]{2}}}}%
{B1m}{\tensor{\chi}{^{\flat}_{\perp}_{\indiq[#1]{1}}}}%
{B2p}{\tensor{\smash{\overset{\sim}{\chi}}}{^{\flat}_{\indiq[#1]{2}}}}%
{A0p}{\tensor{\chi}{^{\flat}_\perp}}%
{A0m}{\tensor[^{\text{P}}]{\mathcal{T}}{^{\flat}}}%
{A1p}{\tensor{\smash{\overset{\wedge}{\chi}}}{^{\flat}_{\perp\indiq[#1]{2}}}}%
{A1m}{\tensor{\smash{\overset{\rightharpoonup}{\mathcal{T}}}}{^{\flat}_{\indiq[#1]{1}}}}%
{A2p}{\tensor{\smash{\overset{\sim}{\chi}}}{^{\flat}_{\perp\indiq[#1]{2}}}}%
{A2m}{\tensor[^{\text{T}}]{\mathcal{T}}{^{\flat}_{\indiq[#1]{3}}}}%
}[\tensor{\mathcal{T}}{^{\flat}_{\indiq[#1]{3}}}]%
}

\newrobustcmd{\cT}[2][placeholder]{%
\IfEqCase{#2}{%
{B1p}{\tensor{\mathcal{T}}{_{\perp\indiq[#1]{2}}}}%
{B1m}{\tensor{\overset{\rightharpoonup}{\mathcal{T}}}{_{\indiq[#1]{1}}}}%
{A0m}{\tensor[^{\text{P}}]{\mathcal{T}}{}}%
{A2m}{\tensor[^{\text{T}}]{\mathcal{T}}{_{\indiq[#1]{3}}}}%
}[\packageError{cosmicclass}{Unidentified Critical Case: #1}{}]%
}

\newrobustcmd{\cTLambda}[2][placeholder]{%
\IfEqCase{#2}{%
{B1p}{\tensor{\zeta}{_{\perp\indiq[#1]{2}}}}%
{B1m}{\tensor{\overset{\rightharpoonup}{\zeta}}{_{\indiq[#1]{1}}}}%
{A0m}{\tensor[^{\text{P}}]{\zeta}{}}%
{A2m}{\tensor[^{\text{T}}]{\zeta}{_{\indiq[#1]{3}}}}%
}[\packageError{cosmicclass}{Unidentified Critical Case: #1}{}]%
}

\newrobustcmd{\cTpic}[2][placeholder]{%
\IfEqCase{#2}{%
{B1p}{\tensor{\phi}{_{\perp\indiq[#1]{2}}}}%
{B1m}{\tensor{\overset{\rightharpoonup}{\phi}}{_{\indiq[#1]{1}}}}%
{A0m}{\tensor[^{\text{P}}]{\phi}{}}%
{A2m}{\tensor[^{\text{T}}]{\phi}{_{\indiq[#1]{3}}}}%
}[\packageError{cosmicclass}{Unidentified Critical Case: #1}{}]%
}

\newrobustcmd{\cTpicl}[2][placeholder]{%
\IfEqCase{#2}{%
{B1p}{\tensor{\phi}{^{\flat}_{\perp\indiq[#1]{2}}}}%
{B1m}{\tensor{\overset{\rightharpoonup}{\phi}}{^{\flat}_{\indiq[#1]{1}}}}%
{A0m}{\tensor[^{\text{P}}]{\phi}{^{\flat}}}%
{A2m}{\tensor[^{\text{T}}]{\phi}{^{\flat}_{\indiq[#1]{3}}}}%
}[\packageError{cosmicclass}{Unidentified Critical Case: #1}{}]%
}

\newrobustcmd{\cTPiP}[2][placeholder]{%
\IfEqCase{#2}{%
{B1p}{\tensor{\varpi}{_{\perp\indiq[#1]{2}}}}%
{B1m}{\tensor{\overset{\rightharpoonup}{\varpi}}{_{\indiq[#1]{1}}}}%
{A0m}{\tensor[^{\text{P}}]{\varpi}{}}%
{A2m}{\tensor[^{\text{T}}]{\varpi}{_{\indiq[#1]{3}}}}%
}[\packageError{cosmicclass}{Unidentified Critical Case: #1}{}]%
}

\newrobustcmd{\cTl}[2][placeholder]{%
\IfEqCase{#2}{%
{B1p}{\tensor{\mathcal{T}}{^{\flat}_{\perp\indiq[#1]{2}}}}%
{B1m}{\tensor{\smash{\overset{\rightharpoonup}{\mathcal{T}}}}{^{\flat}_{\indiq[#1]{1}}}}%
{A0m}{\tensor[^{\text{P}}]{\mathcal{T}}{^{\flat}}}%
{A2m}{\tensor[^{\text{T}}]{\mathcal{T}}{^{\flat}_{\indiq[#1]{3}}}}%
}[\packageError{cosmicclass}{Unidentified Critical Case: #1}{}]%
}

\newrobustcmd{\cTu}[2][placeholder]{%
\IfEqCase{#2}{%
{B1p}{\tensor{\mathcal{T}}{^{\perp\indiq[#1]{2}}}}%
{B1m}{\tensor{\smash{\overset{\rightharpoonup}{\mathcal{T}}}}{^{\indiq[#1]{1}}}}%
{A0m}{\tensor[^{\text{P}}]{\mathcal{T}}{}}%
{A2m}{\tensor[^{\text{T}}]{\mathcal{T}}{^{\indiq[#1]{3}}}}%
}[\packageError{cosmicclass}{Unidentified Critical Case: #1}{}]%
}

\newrobustcmd{\ncTLambda}[2][placeholder]{%
\IfEqCase{#2}{%
{B0p}{\tensor{\zeta}{^{\indiq[#1]{1}}_{\indiq[#1]{1}\perp}}}%
{B1p}{\tensor{\zeta}{_{[\indiq[#1]{2}]\perp}}}%
{B1m}{\tensor{\zeta}{_{\perp\indiq[#1]{1}\perp}}}%
{B2p}{\tensor{\zeta}{_{\langle\indiq[#1]{2}\rangle\perp}}}%
}[\packageError{cosmicclass}{Unidentified Critical Case: #1}{}]%
}

\newrobustcmd{\ncTmul}[2][placeholder]{%
\IfEqCase{#2}{%
{B0p}{\tensor{\upsilon}{^{\indiq[#1]{1}}_{\indiq[#1]{1}\perp}}}%
{B1p}{\tensor{\upsilon}{_{[\indiq[#1]{2}]\perp}}}%
{B1m}{\tensor{\upsilon}{_{\perp\indiq[#1]{1}\perp}}}%
{B2p}{\tensor{\upsilon}{_{\langle\indiq[#1]{2}\rangle\perp}}}%
}[\packageError{cosmicclass}{Unidentified Critical Case: #1}{}]%
}

\newrobustcmd{\ncTpic}[2][placeholder]{%
\IfEqCase{#2}{%
{B0p}{\tensor{\phi}{^{\indiq[#1]{1}}_{\indiq[#1]{1}\perp}}}%
{B1p}{\tensor{\phi}{_{[\indiq[#1]{2}]\perp}}}%
{B1m}{\tensor{\phi}{_{\perp\indiq[#1]{1}\perp}}}%
{B2p}{\tensor{\phi}{_{\langle\indiq[#1]{2}\rangle\perp}}}%
}[\packageError{cosmicclass}{Unidentified Critical Case: #1}{}]%
}

\newrobustcmd{\ncTpicl}[2][placeholder]{%
\IfEqCase{#2}{%
  {B0p}{\tensor{\phi}{^{\flat}^{\indiq[#1]{1}}_{\indiq[#1]{1}\perp}}}%
{B1p}{\tensor{\phi}{^{\flat}_{[\indiq[#1]{2}]\perp}}}%
{B1m}{\tensor{\phi}{^{\flat}_{\perp\indiq[#1]{1}\perp}}}%
{B2p}{\tensor{\phi}{^{\flat}_{\langle\indiq[#1]{2}\rangle\perp}}}%
}[\packageError{cosmicclass}{Unidentified Critical Case: #1}{}]%
}

\newrobustcmd{\ncTPiP}[2][placeholder]{%
\IfEqCase{#2}{%
{B0p}{\tensor{\varpi}{^{\indiq[#1]{1}}_{\indiq[#1]{1}\perp}}}%
{B1p}{\tensor{\varpi}{_{[\indiq[#1]{2}]\perp}}}%
{B1m}{\tensor{\varpi}{_{\perp\indiq[#1]{1}\perp}}}%
{B2p}{\tensor{\varpi}{_{\langle\indiq[#1]{2}\rangle\perp}}}%
}[\packageError{cosmicclass}{Unidentified Critical Case: #1}{}]%
}

\newrobustcmd{\ncT}[2][placeholder]{%
\IfEqCase{#2}{%
{B0p}{\tensor{\mathcal{T}}{^{\indiq[#1]{1}}_{\indiq[#1]{1}\perp}}}%
{B1p}{\tensor{\mathcal{T}}{_{[\indiq[#1]{2}]\perp}}}%
{B1m}{\tensor{\mathcal{T}}{_{\perp\indiq[#1]{1}\perp}}}%
{B2p}{\tensor{\mathcal{T}}{_{\langle\indiq[#1]{2}\rangle\perp}}}%
}[\packageError{cosmicclass}{Unidentified Critical Case: #1}{}]%
}

\newrobustcmd{\cR}[2][placeholder]{%
\IfEqCase{#2}{%
{A0p}{\tensor{\underline{\mathcal{R}}}{}}%
{A0m}{\tensor[^{\text{P}}]{\mathcal{R}}{_{\perp\circ}}}%
{A1p}{\tensor{\underline{\mathcal{R}}}{_{[\indiq[#1]{2}]}}}%
{A1m}{\tensor{\mathcal{R}}{_{\perp\indiq[#1]{1}}}}%
{A2p}{\tensor{\underline{\mathcal{R}}}{_{\langle\indiq[#1]{2}\rangle}}}%
{A2m}{\tensor[^{\text{T}}]{\mathcal{R}}{_{\perp\indiq[#1]{3}}}}%
}[\packageError{cosmicclass}{Unidentified Critical Case: #1}{}]%
}

\newrobustcmd{\cRLambda}[2][placeholder]{%
\IfEqCase{#2}{%
{A0p}{\tensor{\underline{\zeta}}{}}%
{A0m}{\tensor[^{\text{P}}]{\zeta}{_{\perp\circ}}}%
{A1p}{\tensor{\underline{\zeta}}{_{[\indiq[#1]{2}]}}}%
{A1m}{\tensor{\zeta}{_{\perp\indiq[#1]{1}}}}%
{A2p}{\tensor{\underline{\zeta}}{_{\langle\indiq[#1]{2}\rangle}}}%
{A2m}{\tensor[^{\text{T}}]{\zeta}{_{\perp\indiq[#1]{3}}}}%
}[\packageError{cosmicclass}{Unidentified Critical Case: #1}{}]%
}

\newrobustcmd{\cRl}[2][placeholder]{%
\IfEqCase{#2}{%
  {A0p}{\tensor{\underline{\mathcal{R}}}{^{\flat}}}%
{A0m}{\tensor[^{\text{P}}]{\mathcal{R}}{^{\flat}_{\perp\circ}}}%
{A1p}{\tensor{\underline{\mathcal{R}}}{^{\flat}_{[\indiq[#1]{2}]}}}%
{A1m}{\tensor{\mathcal{R}}{^{\flat}_{\perp\indiq[#1]{1}}}}%
{A2p}{\tensor{\underline{\mathcal{R}}}{^{\flat}_{\langle\indiq[#1]{2}\rangle}}}%
{A2m}{\tensor[^{\text{T}}]{\mathcal{R}}{^{\flat}_{\perp\indiq[#1]{3}}}}%
}[\packageError{cosmicclass}{Unidentified Critical Case: #1}{}]%
}

\newrobustcmd{\cRu}[2][placeholder]{%
\IfEqCase{#2}{%
{A0p}{\tensor{\underline{\mathcal{R}}}{}}%
{A0m}{\tensor[^{\text{P}}]{\mathcal{R}}{_{\perp\circ}}}%
{A1p}{\tensor{\underline{\mathcal{R}}}{^{[\indiq[#1]{2}]}}}%
{A1m}{\tensor{\mathcal{R}}{^{\perp\indiq[#1]{1}}}}%
{A2p}{\tensor{\underline{\mathcal{R}}}{^{\langle\indiq[#1]{2}\rangle}}}%
{A2m}{\tensor[^{\text{T}}]{\mathcal{R}}{^{\perp\indiq[#1]{3}}}}%
}[\packageError{cosmicclass}{Unidentified Critical Case: #1}{}]%
}

\newrobustcmd{\ncR}[2][placeholder]{%
\IfEqCase{#2}{%
  {A0p}{\tensor{\mathcal{R}}{_{\perp\perp}}}%
{A0m}{\tensor[^{\text{P}}]{\mathcal{R}}{_{\circ\perp}}}%
{A1p}{\tensor{\mathcal{R}}{_{\perp[\indiq[#1]{2}]\perp}}}%
{A1m}{\tensor{\mathcal{R}}{_{\indiq[#1]{1}\perp}}}%
{A2p}{\tensor{\mathcal{R}}{_{\perp\langle\indiq[#1]{2}\rangle\perp}}}%
{A2m}{\tensor[^{\text{T}}]{\mathcal{R}}{_{\indiq[#1]{3}\perp}}}%
}[\packageError{cosmicclass}{Unidentified Critical Case: #1}{}]%
}

\newrobustcmd{\ncRLambda}[2][placeholder]{%
\IfEqCase{#2}{%
  {A0p}{\tensor{\zeta}{_{\perp\perp}}}%
{A0m}{\tensor[^{\text{P}}]{\zeta}{_{\circ\perp}}}%
{A1p}{\tensor{\zeta}{_{\perp[\indiq[#1]{2}]\perp}}}%
{A1m}{\tensor{\zeta}{_{\indiq[#1]{1}\perp}}}%
{A2p}{\tensor{\zeta}{_{\perp\langle\indiq[#1]{2}\rangle\perp}}}%
{A2m}{\tensor[^{\text{T}}]{\zeta}{_{\indiq[#1]{3}\perp}}}%
}[\packageError{cosmicclass}{Unidentified Critical Case: #1}{}]%
}

\newrobustcmd{\Proj}[2][placeholder]{%
\IfEqCase{#2}{%
  {A2m}{\tensor[^{\text{T}}]{\check{\mathcal{P}}}{#1}}%
}[\packageError{cosmicclass}{Unidentified Critical Case: #1}{}]%
}

\newrobustcmd{\Projl}[2][placeholder]{%
\IfEqCase{#2}{%
  {A2m}{\tensor[^{\text{T}}]{\check{\mathcal{P}}}{^{\flat}#1}}%
}[\packageError{cosmicclass}{Unidentified Critical Case: #1}{}]%
}

\newrobustcmd{\fA}{%
  {\tensor{\mathcal{  A}}{_{\acu{u}}}}%
}
\newrobustcmd{\fB}{%
  {\tensor{\mathcal{  B}}{_{\acu{v}}}}%
}
\newrobustcmd{\fC}{%
  {\tensor{\mathcal{  C}}{^{\acu{v}}}}%
}
\newrobustcmd{\fphi}{%
  {\tensor{\phi}{^{\acu{w}}}}%
}
\newrobustcmd{\fpi}{%
  {\tensor{\pi}{_{\acu{w}}}}%
}

\newrobustcmd{\covard}[2]{%
  {\frac{\bar{\delta}#1}{\bar{\delta}#2}}
}
\newrobustcmd{\copard}[2]{%
  {\frac{\bar{\partial}#1}{\bar{\partial}#2}}
}
\newrobustcmd{\pard}[2]{%
  {\frac{\partial #1}{\partial #2}}
}

\newrobustcmd{\PPM}[1]{%
  {\left[\tensor*{\mathsf{M}}{_{\ }^{\left(\text{#1}\right)}}\right]}%
}

\tikzset{
  good/.style={circle, opacity=0.7, draw=green!60, fill=green!5, line width=.8mm, minimum size=3.5mm},
  bad/.style={circle, opacity=0.7, draw=red!60, fill=red!5, line width=.8mm, minimum size=3.5mm},
  badx/.style={circle, opacity=0.4, draw=red!60, fill=red!5, line width=.8mm, minimum size=3.5mm},
  save/.style={circle, opacity=0.7, draw=green!60, fill=green!60, line width=1.5mm, minimum size=3.5mm},
  kill/.style={circle, opacity=0.7, draw=red!60, fill=red!60, line width=1.5mm, minimum size=3.5mm},
  bind/.style={draw=green!60, opacity=0.7, line width=1mm,},
  wrap/.style={draw=red!60, opacity=0.7, line width=1mm,},
}

\DeclareTOCStyleEntry[numwidth=20pt,linefill=\bfseries\TOCLineLeaderFill]{tocline}{section}
\DeclareTOCStyleEntry[entryformat=\textit,numwidth=10pt,linefill=\TOCLineLeaderFill]{tocline}{subsection}
\DeclareTOCStyleEntry[entryformat=\textit,numwidth=10pt,linefill=\TOCLineLeaderFill]{tocline}{subsubsection}

\usepackage{hyperref}
\hypersetup{     colorlinks = true,     linkcolor = Blue,     citecolor = Blue,     filecolor = Blue,     urlcolor = Blue%
     }\usepackage[capitalize]{cleveref}

\begin{document}

\title{Buchdahl bound, photon ring, ISCO and radial acceleration in Einstein-\ae{}ther theory}

\author{Yi-Hsiung Hsu}
\email{yhh36@cam.ac.uk}
\affiliation{Astrophysics Group, Cavendish Laboratory, JJ Thomson Avenue, Cambridge CB3 0HE, UK}

\author{Anthony Lasenby}
\email{a.n.lasenby@mrao.cam.ac.uk}
\affiliation{Astrophysics Group, Cavendish Laboratory, JJ Thomson Avenue, Cambridge CB3 0HE, UK}
\affiliation{Kavli Institute for Cosmology, Madingley Road, Cambridge CB3 0HA, UK}

\author{Will Barker}
\email{wb263@cam.ac.uk}
\affiliation{Astrophysics Group, Cavendish Laboratory, JJ Thomson Avenue, Cambridge CB3 0HE, UK}
\affiliation{Kavli Institute for Cosmology, Madingley Road, Cambridge CB3 0HA, UK}
\affiliation{Central European Institute for Cosmolgy and Fundamental Physics, Institute of Physics of the Czech Academy of Sciences, Na Slovance 1999/2, 182 00 Prague 8, Czechia}

\author{Amel Durakovic}
\email{amel@fzu.cz}
\affiliation{Observatoire astronomique de Strasbourg, Université de Strasbourg, 11 Rue de l’Université, 67000 Strasbourg, France}
\affiliation{Central European Institute for Cosmolgy and Fundamental Physics, Institute of Physics of the Czech Academy of Sciences, Na Slovance 1999/2, 182 00 Prague 8, Czechia}

\author{Michael Hobson}
\email{mph@mrao.cam.ac.uk}
\affiliation{Astrophysics Group, Cavendish Laboratory, JJ Thomson Avenue, Cambridge CB3 0HE, UK}

\begin{abstract}
	Spherically symmetric Einstein-{\ae}ther (E{\AE}) theory with a Maxwell-like kinetic term is revisited. We consider a general choice of the metric and the \ae{}ther field, finding that:~(i) there is a gauge freedom allowing one always to use a diagonal metric; and~(ii) the nature of the Maxwell equation forces the \ae{}ther field to be time-like in the coordinate basis. We derive the vacuum solution and confirm that the innermost stable circular orbit (ISCO) and photon ring are enlarged relative to general relativity (GR). Buchdahl's theorem in E\AE{} theory is derived. For a uniform physical density, we find that the upper bound on compactness is always lower than in GR. Additionally, we observe that the Newtonian and E\AE{} radial acceleration relations run parallel in the low pressure limit. Our analysis of E\AE{} theory may offer novel insights into its interesting phenomenological generalization: \AE{}ther--scalar--tensor theory ({\AE}ST).
\end{abstract}
\maketitle
\tableofcontents

\section{Introduction}\label{sec:intro}
General relativity (GR) has been a successful theory for describing gravity; however, interest in going beyond the conventional gravity theory has never ceased.  \AE{}ther--scalar--tensor (\AE{}ST) theory~\cite{Skordis:2020eui,Bataki:2023uuy,Skordis:2021mry,Verwayen:2023sds,Llinares:2023lky,Mistele:2023paq,Mistele:2023fwd,Bernardo:2022acn,Tian:2023gjt,Kashfi:2022dyb,Rosa:2023qun} is a promising modified gravity theory, which is fully relativistic but can nevertheless lead to behaviour similar to modified Newtonian dynamics (MOND) on intermediate scales~\cite{Verwayen:2023sds,Llinares:2023lky,Mistele:2023paq}, whilst being compatible with CDM on the largest scales~\cite{Tian:2023gjt,Kashfi:2022dyb,Rosa:2023qun}  and having tensor gravitational waves (GWs) which propagate at the speed of light~\cite{Skordis:2020eui,Skordis:2021mry,Bataki:2023uuy}. This is in contrast to the hitherto most successful candidate for a relativistic theory incorporating MOND effects, namely, tensor-vector-scalar (TeVeS) theory~\cite{Bekenstein:1984tv,Bekenstein:1988zy,Sanders:1996wk}, which is phenomenologically inconsistent with the observed GWs~\cite{LIGOScientific:2016aoc,2018IJMPD..2747027S,Skordis:2019fxt} and cosmic microwave background (CMB)~\cite{Planck:2018vyg,Planck:2015fie,Planck:2013pxb}. \AE{}ST holds out of the hope of being able to provide an explanation in terms of modified gravity for some of the regularities in the radial acceleration relations for galaxies, for which there is currently strong evidence in the data~\cite{Milgrom:1983zz,Milgrom:1983ca,Bekenstein:1984tv,Bekenstein:1988zy,Sanders:1996wk,Bekenstein:2004ne,Skordis:2009bf} (see also reviews~\cite{Milgrom:2001ny,Famaey:2011kh}). Moreover, it achieves this in a fully relativistic context. The Lagrangian and equations of motion for the theory are rather complicated, however, and it is of interest to investigate solutions both for matter and vacuum scenarios in a simpler context which nevertheless captures some of the essential features of \AE{}ST.

Such a theory is the much older Einstein--\ae ther (E\AE) theory, relative to which  \AE{}ST theory can be viewed as an extension with an extra scalar field. In 1951 Dirac briefly considered an alternative to the then-nascent (or ``\textit{ugly and incomplete}''~\cite{Dirac:1951}) renormalisation of electron self-energy~\cite{Bethe:1947id,Tomonaga:1946zz,Schwinger:1948iu,Feynman:1949zx,Feynman:1949hz,Dyson:1949bp,Dyson:1949ha}. In particular, Dirac proposed the normalised electromagnetic potential to be unit-timelike on shell 
\begin{equation}\label{UnitTimelike}
	\tensor{A}{_\mu}\tensor{A}{^\mu}\onshell 1.
\end{equation}
Today we instead imagine such~$\tensor{A}{^\mu}$ as being the four-velocity of an \ae ther fluid~\cite{Balakin:2020vmn}. In contrast~\cite{Michelson:1887zz,Navarro:2021} to its \textit{luminiferous} \ae ther namesake, this \ae ther is not a fixed, non-dynamical medium, but a dynamical field that can evolve and interact with gravity and matter. Fields which obey~\cref{UnitTimelike} moreover have a growing importance in modified theories of gravity~\cite{Bluhm:2008yt,Bluhm:2007bd,PhysRevD.84.044056,Blanchet:2024mvy}. By embedding Dirac's model directly into GR, we obtain the minimal E\AE{} theory with a Maxwell-like kinetic term~\cite{Clifton:2011jh,Jacobson:2000xp,Eling:2003rd,Jacobson:2004ts,Eling:2004dk,Foster:2005dk,Eling:2006ec,Konoplya:2006rv,Eling:2007xh,Jacobson:2007veq,Garfinkle:2007bk,Tamaki:2007kz,Barausse:2011pu,Berglund:2012fk,Berglund:2012bu,Gao:2013im,Barausse:2015frm,Campista:2018gfi,Bhattacharjee:2018nus,Oost:2018tcv,Lin:2018ken,Zhang:2019iim,Zhu:2019ura,Leon:2019jnu,Coley:2019tyx,Zhang:2020too,Adam:2021vsk,Chan:2022mxd} (see the review~\cite{Monfort-Urkizu:2023jop}), which is an important progenitor to other modern theories of modified gravity, such as bumblebee~\cite{Kostelecky:1989jw,Colladay:1998fq,Kostelecky:2003fs} and Ho\v{r}ava gravity~\cite{Jacobson:2013xta,Horava:2009uw,Visser:2011mf,Herrero-Valea:2023zex}, as well as to \AE ST.

The action of the minimal E\AE{} theory can be written as
\begin{align}\label{EAeLagrangian}
	S_{\text{E\AE{}}}\equiv\int\mathrm{d}^4x\sqrt{-g}\bigg[&-\frac{\Planck^2}{2}R-\Kb\Planck^2\tensor{F}{_{\mu\nu}}\tensor{F}{^{\mu\nu}}
	\nonumber\\
	&\hspace{-10pt}
	-2\lambda\left(\tensor{A}{_\mu}\tensor{A}{^\mu}-1\right)-\Planck^2\Lambda+L_{\text{M}}\bigg],
\end{align}
which is fully parameterised by the Planck mass~$\Planck$ and cosmological constant~$\Lambda$ from GR, and also the dimensionless E\AE{} coupling~$\Kb$ which lies in the range
\begin{equation}\label{KbRange}
	0<\Kb<2.
\end{equation}
In~\cref{EAeLagrangian}~$\tensor{F}{_{\mu\nu}}\equiv 2\tensor{\nabla}{_{[\mu}}\tensor{A}{_{\nu]}}$, whilst~$R\equiv\tensor*{R}{^\mu_\mu}\equiv\tensor{R}{_{\mu\nu}^{\mu\nu}}$ is the Ricci scalar, and~$\lambda$ is a Lagrange multiplier enforcing~\cref{UnitTimelike}. In~$L_{\text{M}}$ we put all other matter fields, including standard electromagnetism --- though a large literature (see e.g.~\cite{Dymnikova:2023,Dymnikova:2015hka,Dymnikova:2015hka,Dymnikova:2021}) now motivates E\AE{} extensions to~$f(F)$ or~$\tensor{A}{_\mu}\tensor{A}{_\nu}\tensor{R}{^{\mu\nu}}$ and~$\tensor{A}{_\mu}\tensor{A}{^\mu}\tensor{R}{}$ operators, or `soft' realisations of~\cref{UnitTimelike} at potential vacua, via effective nonlinear quantum electrodynamics (QED). 

From the equation of motion with respect to~$A^\mu$ derived from~\cref{EAeLagrangian}, one finds that 
\begin{align}
\lambda = -\Kb \Planck^2 \tensor{A}{^\mu} \tensor{\nabla}{^\nu} \tensor{F}{_{\mu\nu}}.
\end{align}
Therefore, the non-GR part in~\cref{EAeLagrangian} is all proportional to~$\Kb$, which means~$\Kb \mapsto0$ corresponds to the GR limit. The need for limited range of~$\Kb$ values in ~\cref{KbRange} is not at all obvious from~\cref{EAeLagrangian}, but it does emerge in the E\AE{} field equations, which not only yield exact GR equivalence for~${\Kb\mapsto 0}$, but also identify ~${\Kb\mapsto 2}$ with the `extremal \ae ther' regime~\cite{Eling:2004dk}. 

In this paper, we focus on vacuum and matter solutions of minimal E\AE{} theory under the assumption of spherical symmetry. In particular, we derive Buchdahl's theorem in E\AE{} theory; this sets an absolute bound on the compactness of spherically symmetric objects in the theory, which may be relevant to the physics of neutron stars. Additionally, we investigate the radial acceleration curve, i.e., the acceleration that would be observed in E\AE{} theory compared to what would be seen in Newtonian gravity. While E\AE{} theory is not considered a viable candidate for replacement of GR, it is plausible that some of the properties we consider will also be found in the more promising \AE{}ST theory, although this remains a topic for further research.

The structure of this paper is as follows. In~\cref{sec:EAtheory}, we first examine the spherically symmetric vacuum solution and its associated gauge invariance, followed in~\cref{sec:matterin} by a discussion of the solution with matter content. In particular, we derive Buchdahl's theorem within the context of E\AE{} theory. In~\cref{sec:GC}, we present the scaling relation of the radial acceleration relation, comparing the E\AE{} profiles to the Newtonian and MOND ones. Conclusions follow in~\cref{sec:conclusion}. Throughout this paper, we adopt geometrised units, setting~$G\equiv c\equiv 1$, unless otherwise specified.

\section{The vacuum exterior}\label{sec:EAtheory}

The general form of the spherically symmetric line element is discussed using the tetrad formalism in~\cite{Kim:2018aa}, and originates from the functions shown in Table IV of~\cite{Lasenby:1998yq}. In a spacetime labelled with some set of coordinates~$\tensor{x}{^\mu}$, the (holonomic) coordinate basis vectors~$\tensor{\mathbf{e}}{_\mu}$ (denoted by Greek indices) at each point are related to the metric via~$\tensor{\mathbf{e}}{_\mu}\cdot\tensor{\mathbf{e}}{_\nu} \equiv \tensor{g}{_{\mu\nu}}$. At each point one may also define a local Lorentz frame by another set of (anholonomic) orthonormal basis vectors~$\tensor{\hat{\mathbf{e}}}{_a}$
 (denoted by Roman indices), for which~$\tensor{\hat{\mathbf{e}}}{_a} \cdot \tensor{\hat{\mathbf{e}}}{_a} \equiv \tensor{\eta}{_{ab}}$, where~$\tensor{\eta}{_{ab}} \equiv \mbox{diag}(1, -1, -1, -1)$ is the Minkowski metric. The two sets of basis vectors are related by the tetrads (or vierbeins)~$\tensor{e}{_a^\mu}$, where the inverse is denoted~$\tensor{e}{^a_\mu}$, such that~$\tensor{\hat{\mathbf{e}}}{_a} \equiv \tensor{e}{_a^\mu} \tensor{\mathbf{e}}{_\mu}$ and~$\tensor{\mathbf{e}}{_\mu} \equiv \tensor{e}{^a_\mu} \tensor{\hat{\mathbf{e}}}{_a}$. It is straightforward to show that~$\tensor{g}{_{\mu\nu}} \equiv \tensor{\eta}{_{ab}}\tensor{e}{^a_\mu}\tensor{e}{^b_\nu}$, which is invariant under local rotations of the Lorentz frames.

For a stationary, spherically-symmetric system and assuming a spherical polar coordinate system~$[\tensor{x}{^\mu}] = (t,r,\theta,\phi)$, one may choose the only non-zero tetrad coefficients to be~$\tensor{e}{_0^0} \equiv f_1(r)$,~$\tensor{e}{_1^0} \equiv f_2(r)$,~$\tensor{e}{_1^1} \equiv g_1(r)$,~$\tensor{e}{_0^1} \equiv g_2(r)$,~$\tensor{e}{_2^2} \equiv 1/r$ and~$\tensor{e}{_3^3} = 1/(r\sin\theta)$.  In so doing, we have made use of the invariance under local rotations of the Lorentz frames to align~$\tensor{\hat{\mathbf{e}}}{_2}$ and 
$\tensor{\hat{\mathbf{e}}}{_3}$ with
the coordinate basis vectors~$\tensor{\mathbf{e}}{_2}$ and 
$\tensor{\mathbf{e}}{_3}$ at each point. As discussed in ~\cite{Lasenby:1998yq}, it is also convenient to adopt the `Newtonian gauge', in which~$f_2 = 0$. One may then
write the line-element as
\begin{align}
    \mathrm{d} s^2=\frac{g_1^2-g_2^2}{f_1^2 g_1^2} \, \mathrm{d} t^2&+\frac{2 g_2}{f_1 g_1^2} \, \mathrm{d} t \,\mathrm{d} r-\frac{1}{g_1^2} \, \mathrm{d} r^2\nonumber\\
    &-r^2 \left(\mathrm{d}\theta^2+\sin^2\theta \,\mathrm{d}\phi^2\right),
\label{eq:general_line_element}
\end{align}
which involves three scalar functions of~$r$, where we have adopted a ‘physical’ (non-comoving) radial coordinate for which the proper area of a sphere of radius~$r$ is~$4\pi r^2$. The line-element in~\cref{eq:general_line_element} possesses a single further gauge freedom, which is the direction of the timelike unit frame vector~$\tensor{\hat{\mathbf{e}}}{_0} = f_1 \tensor{\mathbf{e}}{_0} + g_2 \tensor{\mathbf{e}}{_1}$ at each point. This may be chosen to coincide with the four-velocity of some radially-moving test particle or observer (which need not be in free-fall), so that the components of its 4-velocity in the tetrad frame are~$[\tensor{u}{^a}] = [1,0,0,0]$, and hence in the coordinate basis one has~$[\tensor{u}{^\mu}]= [\dot{t},\dot{r},\dot{\theta},\dot{\phi}] = [f_1,g_2,0,0]$, where dots denote differentiation with respect to the observer’s proper time. In the presence of a fluid, it is most natural to choose the observer to be comoving with the fluid; indeed, depending on the gravitational theory and physical system under consideration, this coincidence may be required by the equations of motion. In any case, since~$g_2$
 is the rate of change of the~$r$ coordinate of the observer with respect to its proper time, it can be physically interpreted as the observer's three-velocity, which is consistent with its presence in the~$\mathrm{d}t \, \mathrm{d}r$ cross term of~\cref{eq:general_line_element}.
 
\subsection{The Eling--Jacobson solution}\label{sec:Jacobson}
In this section, we obtain fresh details about the central anatomy of the asymptotically flat vacuum solution to the E\AE{} theory in~\cref{EAeLagrangian}. This solution was first identified in~\cite{Eling:2006df} by aligning the \ae{}ther field with the timelike Killing vector. It shares the Newtonian limit of the Schwarzschild black hole in GR, and indeed the solutions are fully identical in the~$\Kb\mapsto 0$ limit. For finite~$\Kb$, however, the innermost stable circular orbit (ISCO), photon ring and singular surface of the Schwarzschild-like coordinates all lie at \textit{increased radii}. The derivation of the vacuum solution is provided here, serving as a foundation for the subsequent discussions on the ISCO in~\cref{sec:ISCO}, mass conversion in~\cref{sec:buchdahl}, and boundary matching in~\cref{sec:appendix}.

We look for static, Schwarzschild-like vacuum solutions to~\cref{EAeLagrangian} with vanishing vacuum energy~$\Lambda=0$. We thus choose~$g_2=0$, so that 
\begin{align}\label{Schwarzschild}
	\mathrm{d}s^2\onshell \TimeFunc \mathrm{d} t^2 - \SpaceFunc \mathrm{d} r^2 - r^2 \left(\mathrm{d}\theta^2 + \sin^2 \theta\mathrm{d} \phi^2  \right),
\end{align}
where~$\TimeFunc\equiv\TimeFunc(r)=1/f^2_1(r)$ and~$\SpaceFunc\equiv\SpaceFunc(r)   = 1/g_1^2(r)$ are dimensionless functions to be determined. We want to find the closest analogue to the Schwarzschild exterior solution in GR
\begin{equation}\label{SchwarzschildActual}
	\TimeFunc(r)\onshell\frac{1}{\SpaceFunc(r)}\onshell 1-\frac{\Rg}{r},\quad \Rg<r,
\end{equation}
which is fully parameterised by the constant Schwarzschild radius~${\Rg\equiv -2\Pg r}$, where~$\Pg$ is the Newtonian potential associated with the (gravitational) mass of the interior. Guided by~\cref{SchwarzschildActual}, we will focus on 
\begin{equation}\label{SchwarzschildRange}
	0<\TimeFunc\leq 1\leq\SpaceFunc.
\end{equation}
We further assume complete alignment of the \ae ther with the timelike Killing vector, so from~\cref{UnitTimelike,Schwarzschild}
\begin{equation}\label{Killing}
	A^t(r) \onshell 1/\sqrt{\TimeFunc (r)},\quad A^r(r) \onshell 0.
\end{equation}
Whilst~\cref{Killing} seems somewhat arbitrary, it can be shown that a vanishing radial component is demanded by the Maxwell equation, which is discussed in more detail in~\cref{sec:GI}. Moreover,~\cref{Killing} does yield an isotropic spacetime at infinite radius, and is hence consistent with the cosmological principle. Combining~\cref{Schwarzschild,Killing} the field equations read 
\begin{subequations}
\begin{align}
	0 \onshell &8\left( 1- \SpaceFunc\right) + r \TimeFunc'\TimeFunc^{-2} \left( 8 \TimeFunc + \Kb r \TimeFunc' \right), \label{rrEquation}
	\\
	0 \onshell &3 \Kb \TimeFunc^{-1}\SpaceFunc \TimeFunc'^2 + 2\Kb  r^{-1} \TimeFunc' \left( r \SpaceFunc' -4 \SpaceFunc\right)
	\nonumber\\
	&\ \ \ + 8 r^{-2} \TimeFunc \left[ \left(\SpaceFunc-1 \right) \SpaceFunc + r \SpaceFunc' \right] - 4 \Kb \SpaceFunc \TimeFunc'' , \label{ttEquation}
\end{align}
\end{subequations}
where a prime denotes~$\mathrm{d}/\mathrm{d}r$. It is straightforward to show that~\cref{rrEquation,ttEquation} do indeed admit at spatial infinity a series solution
\begin{subequations}
	\begin{align}
		\TimeFunc(r)&\onshell 1-\frac{\Rg}{r}-\frac{\Kb\Rg^3}{48r^3}+\mathcal{O}\left({\Rg^4}/{r^4}\right),\label{SeriesT}\\
		\SpaceFunc(r)&\onshell 1+\frac{\Rg}{r}+\left(1+\frac{\Kb}{8}\right)\frac{\Rg^2}{r^2}+\mathcal{O}\left({\Rg^3}/{r^3}\right),\label{SeriesS}
	\end{align}
\end{subequations}
which is consistent with the Newtonian limit of~\cref{SchwarzschildActual}, but which departs from the nonlinear Schwarzschild physics for generic E\AE{} parameter~$\Kb$.

It is possible to make~\cref{SeriesT,SeriesS} exact.
We can use~\cref{rrEquation} to determine~$\SpaceFunc$ algebraically, and substitute it into~\cref{ttEquation} to yield
\begin{align}\label{HardEqn}
\Kb r^2 \TimeFunc'^3 + 8 \TimeFunc^2 \left( 2 \TimeFunc' + r \TimeFunc''\right) \onshell 0,
\end{align}
which neatly separates the novel contribution --- the first term in~\cref{HardEqn} --- from what is present in GR to give~\cref{SchwarzschildActual}, i.e.~$2 \TimeFunc'+r\TimeFunc''\onshell 0$.
The nonlinearity of~\cref{HardEqn} makes the E\AE{} solution harder to obtain than the Schwarzschild counterpart. Instead, we obtain a formal expression\footnote{The solution branch (i.e. the sign of the square root) and limits of integration are determined by the Newtonian limit in~\cref{SeriesT,SeriesS}. The discussion of branch choice is demonstrated in~\cref{sec:appendix}.} for~$\TimeFunc$ directly from~\cref{rrEquation}
\begin{align}\label{TIntegral}
	&\TimeFunc\left(r\right) \onshell \nonumber\\
	&\ \ \ \ \ \ \exp\left[ \frac{2}{\Kb}\int^{\infty}_r \mathrm{d}r' \frac{2-\sqrt{4 + 2\Kb \left(\SpaceFunc \left(r'\right)-1\right) }}{r'} \right],
\end{align}
and when~\cref{TIntegral} is substituted into~\cref{ttEquation} with the conditions in~\cref{KbRange,SchwarzschildRange} we find
\begin{align}
	&4\SpaceFunc - 2\SpaceFunc \sqrt{4 + 2\Kb \left(\SpaceFunc-1\right)} \nonumber
	\\
	&\hspace{50pt} + \Kb \left[ 2 \left( \SpaceFunc-1 \right) \SpaceFunc + r \SpaceFunc' \right] \onshell 0,\label{GradEquation}
\end{align}
which integrates to give~$r$ as an inverse function of~$\SpaceFunc$
\begin{align}\label{InverseFunction}
	r\left(\SpaceFunc\right)\onshell & \frac{\Rh\sqrt{2\Kb\SpaceFunc}}{\sqrt{4+2\Kb\left(\SpaceFunc-1\right)}-2}
	\nonumber\\
	&\ \times\left[\frac{\sqrt{2+\Kb\left(\SpaceFunc-1\right)}-\sqrt{2-\Kb}}{\sqrt{2+\Kb\left(\SpaceFunc-1\right)}+\sqrt{2-\Kb}}\right]^{\frac{1}{\sqrt{4-2\Kb}}}.
\end{align}
The integration constant~$\Rh>0$ in~\cref{InverseFunction} has been normalised so as to represent the \emph{throat} radius. As with the Schwarzschild case in~\cref{SchwarzschildActual}, the throat is defined by~$\SpaceFunc\mapsto\infty$, at which point~\cref{InverseFunction} evaluates to\footnote{It is important to observe that, unlike for the Schwarzschild case, the time function~$\TimeFunc(r)$ does not vanish as~$r\mapsto \Rh$. For this reason, we refer to the feature here as a `throat' rather than a `horizon'. It is shown in~\cite{Eling:2006df} that the Eling--Jacobson solution is, in fact, an asymmetric wormhole whose throat is located at~$r=\Rh$. The wormhole nature of the solution may be seen by tracing radial null geodesics, which display a minimum radius here. Except for the discussion in~\cref{sec:appendix} we will avoid considering the `far side' of the wormhole; nor do we focus on the physical nature of the throat surface.}
\begin{equation}\label{HorizonLimit}
	\lim_{\SpaceFunc\mapsto\infty}r\left(\SpaceFunc\right)\onshell \Rh.
\end{equation}
To interpret~\cref{HorizonLimit}, we must fix the other (Newtonian) limit of~\cref{InverseFunction} to that stipulated in~\cref{SeriesS}, giving 
\begin{equation}\label{NewtonianLimit}
	\lim_{\SpaceFunc\mapsto 1}\left(\SpaceFunc-1\right)r\left(\SpaceFunc\right)\onshell \Rg.
\end{equation}
Substituting~\cref{InverseFunction} into~\cref{NewtonianLimit} and evaluating this limit, we determine the necessary relation~$\Rh\equiv \Rh\left(\Rg,\Kb\right)$ to be
\begin{equation}\label{HorizonFunction}
	\Rh\left(\Rg,\Kb\right)\equiv\frac{\Rg\sqrt{\Kb}}{2\sqrt{2}}\left[\frac{\sqrt{2}+\sqrt{2-\Kb}}{\sqrt{2}-\sqrt{2-\Kb}}\right]^{\frac{1}{\sqrt{4-2\Kb}}}.
\end{equation}
We may notice in~\cref{HorizonFunction} the limiting representation of Euler's number~$e$. Recall that the Newtonian limit is associated with a concrete scale~$\Rg$, and~$\Rg$ is also associated with the horizon radius of the Schwarzschild black hole in~\cref{SchwarzschildActual}. But according to~\cref{HorizonFunction}, an E\AE{} vacuum with this same Newtonian limit instead has a throat at~$\Rh$, where 
\begin{equation}\label{HorizonRange}
	\Rg<\Rh<\frac{e}{2}\Rg.
\end{equation}
Thus, for constant gravitational mass, the inner region of the E\AE{} solution is smoothly \textit{enhanced} by a factor of up to~${e/2\simeq 1.359}$ as the E\AE{} parameter~$\Kb$ increases from the GR limit --- through~\cref{KbRange} --- to the `extremal \ae ther' limit. 

In summary, given a Newtonian limit~$\Rg$ to a compact object in the~$\Kb$-specific E\AE{} theory of~\cref{EAeLagrangian}, the analytic formula for the augmented throat radius is given in~\cref{HorizonFunction}, whilst for~$\Rh<r$ the line element functions in~\cref{Schwarzschild} can be obtained from the inverse of the analytic function in~\cref{InverseFunction} and the consequent definite integral in~\cref{TIntegral}.

\subsection{The ISCO and the photon ring}\label{sec:ISCO}

The ISCO of the Eling--Jacobson solution was considered for the first time in~\cite{Eling:2007xh} (the phenomenological implications are discussed in~\cite{Jacobson:2007veq,Jacobson:2007fh}, see also~\cite{Tamaki:2007kz,Barausse:2011pu,Zhang:2020too,Wang:2022yvi} for the ISCOs and~\cite{Wang:2022yvi,Darvishi:2024ndu} for the photon rings around black hole solutions in E\AE{} theory). The stationary observer at~$r>\Rh$ perceives massive particles on circular orbits of radius~$r$ to have orbital speed~$v$ as they pass by, where~$v$ is given by a well-known formula for any Schwarzschild-like line element of the form in~\cref{Schwarzschild}, namely
\begin{equation}\label{VelocityFunction}
	v^2\equiv\frac{r\TimeFunc'}{2\TimeFunc}.
\end{equation}
Substituting~\cref{TIntegral} into~\cref{VelocityFunction} yields
\begin{equation}\label{SpaceInVelocity}
	\SpaceFunc(v)\onshell 1+2v^2+\frac{\Kb}{2}v^4.
\end{equation}
Note that~\cref{SpaceInVelocity} is exact, so we can clearly identify a quartic \ae ther correction proportional to the E\AE{} parameter~$\Kb$.
At large distances, we can verify in~\cref{SeriesS,SpaceInVelocity} the Newtonian result that~$\Rg/2r\equiv-\Pg\onshell v^2+\mathcal{O}\left({\Rg^2}/{r^2}\right)$. More generally,~\cref{SpaceInVelocity} may be substituted into~\cref{InverseFunction} to give~$r$ as an inverse function of~$v$. In this nonlinear regime, we are especially interested in the luminal orbit radius~$\Rl$ for which~$v=1$, since the monotonicity of~\cref{SpaceInVelocity,InverseFunction} guarantees that no massive orbits (even unstable ones) may lie at or within~$\Rl$. We find
\begin{align}\label{Luminal}
	\Rl\left(\Rh,\Kb\right)\equiv & \frac{\Rh}{\sqrt{\Kb}}\sqrt{6+\Kb}
	\nonumber\\
	&\hspace{-0pt} \times\left[\frac{2+\Kb-\sqrt{2}\sqrt{2-\Kb}}{2+\Kb+\sqrt{2}\sqrt{2-\Kb}}\right]^{\frac{1}{\sqrt{4-2\Kb}}},
\end{align}
and by substituting~\cref{HorizonFunction} into~\cref{Luminal} we obtain
\begin{equation}\label{LuminalRange}
	\frac{3}{2}\Rg<\Rl<\sqrt{e}\Rg,
\end{equation}
where the lower bound is consistent with the Schwarzschild case in~\cref{SchwarzschildActual}, as expected. Whilst~$\Rl$ itself is a forbidden orbital radius for massive particles of finite energy, it is equivalently the location of the \textit{only} circular null geodesic. Equatorial null geodesics at~$\uptheta=\pi/2$ in the Schwarzschild-like line element of~\cref{Schwarzschild} are associated with temporal~$k$ and azimuthal~$h$ integrals of motion
\begin{equation}\label{Integrals}
	\left(\frac{\mathrm{d}r}{\mathrm{d}\tau}\right)^2=\frac{k^2}{\TimeFunc\SpaceFunc}-\frac{h^2}{r^2\SpaceFunc},\quad
	k\equiv\TimeFunc\frac{\mathrm{d}t}{\mathrm{d}\tau}, \quad h\equiv r^2\frac{\mathrm{d}\upphi}{\mathrm{d}\tau},
\end{equation}
where~$\tau$ is the affine parameter. By definition,~\cref{Integrals} associates the circular null geodesic with the specific ratio~$\Kl^2/\Hl^2=\TimeFunc(\Rl)/\Rl^2$. By differentiating~\cref{Integrals} with respect to~$\tau$, and applying~\cref{TIntegral,GradEquation,SpaceInVelocity} we obtain near~$\Rl$ an unstable radial oscillator for adjacent (non-circular) null geodesics
\begin{equation}\label{UnstableOscillator}
	\frac{\mathrm{d}^2r}{\mathrm{d}\tau^2}=\frac{\Hl^2}{\Rl^4}\left(r-\Rl\right)+\mathcal{O}\left[\left(r-\Rl\right)^2\right].
\end{equation}
From~\cref{UnstableOscillator} we conclude that the instability of the photon ring which forms at~$\Rl$ is not improved in E\AE{} theory.

\begin{figure}[ht!]
\includegraphics[width=0.5\textwidth]{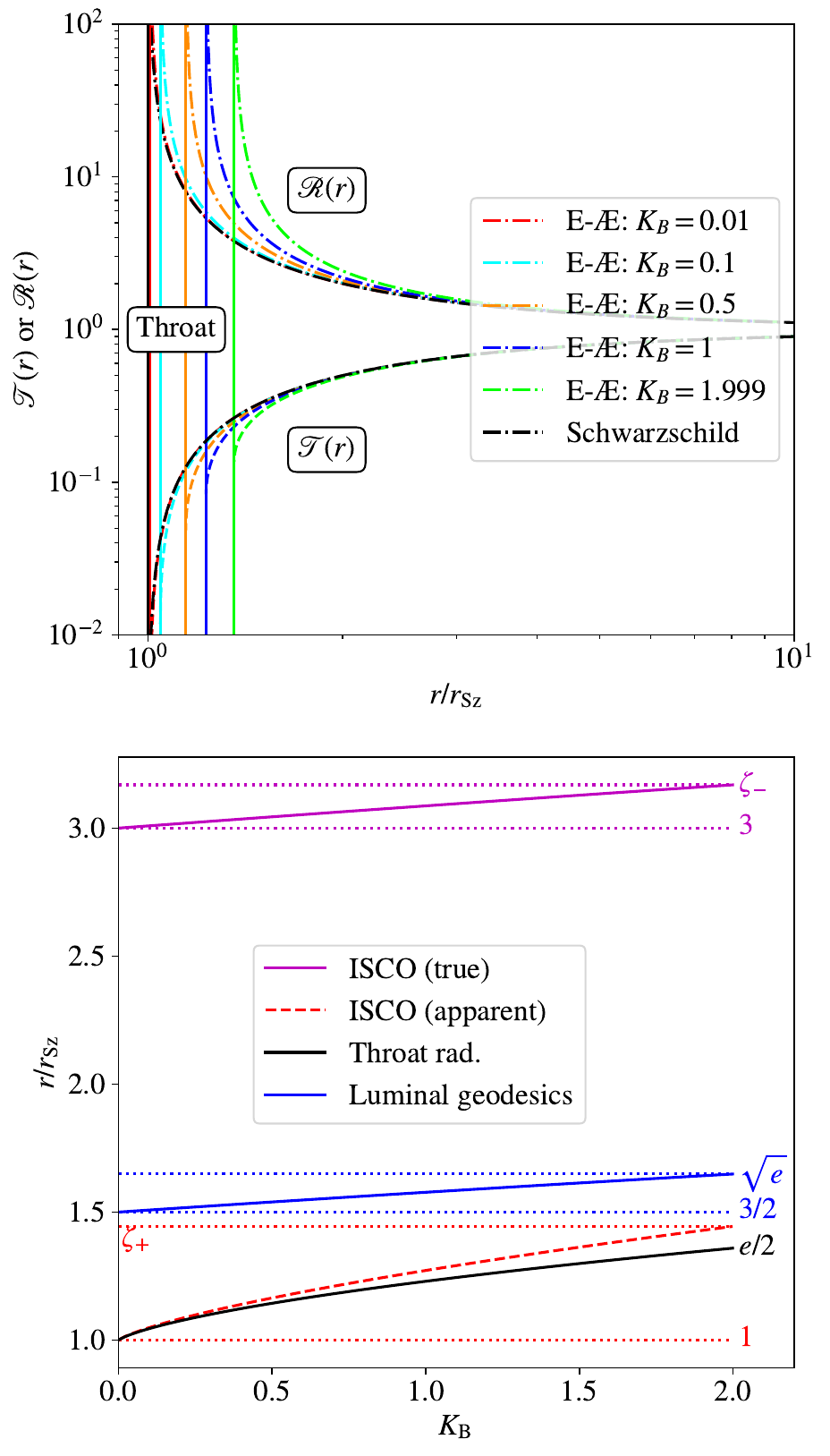}
	\caption{Upper panel: the~$\TimeFunc$ and~$\SpaceFunc$ functions in~\cref{Schwarzschild} for various~$\Kb$ and fixed~$\Rg$, according to the exact E\AE{} exterior solution in~\cref{TIntegral,InverseFunction,HorizonFunction}. Lower panel: the central structures comprise the throat~$\Rh$ in~\cref{HorizonFunction}, the ISCO-like radii~$\Risco$ in~\cref{ISCO}, and the photon ring~$\Rl$ in~\cref{Luminal}. The dimensionless numbers~$\zeta_{\pm}$ correspond (in units of~$\Rg$) to the upper bounds of the ISCO-like radii in~\cref{ISCORange}. These radii all connect with their counterparts for the Schwarzschild black hole of GR in the~$\Kb\mapsto 0$ limit.}
\label{AllPlots}
\end{figure}

We can also consider the stability of the massive orbits. In the massive case~$\tau$ is most conveniently defined as the proper time along the geodesic, whereupon~$k$ is identified with the specific energy and~$h$ with the specific angular momentum. A family of massive circular orbits is possible, and the conditions~$\mathrm{d}r/\mathrm{d}\tau=\mathrm{d}^2r/\mathrm{d}\tau^2=0$ lead --- as with~\cref{VelocityFunction} --- to further results at given~$r$ in any line element of the form in~\cref{Schwarzschild}
\begin{equation}\label{OrbitalEnergy}
	k^2=\frac{2\TimeFunc^2}{2\TimeFunc-r\TimeFunc'},\quad
	h^2=\frac{r^3\TimeFunc'}{2\TimeFunc-r\TimeFunc}.
\end{equation}
Note again that massive orbits will only be possible for real~$k$ or~$h$, i.e., at radii~$r<\Rl$ according to~\cref{OrbitalEnergy,TIntegral,SpaceInVelocity}.
Analogously to~\cref{UnstableOscillator}, it can be shown that the stability criterion for massive circular orbits is~$k'>0$. Substituting~\cref{TIntegral} into~\cref{OrbitalEnergy} and eliminating~$\SpaceFunc'$ with~\cref{GradEquation}, we obtain an algebraic solution for~$k'$ in~$\SpaceFunc$. There are two analytic roots~$k'=0$ lying in the finite exterior, and these are associated with the line element values~$\SpaceFunc_{\pm}$
\begin{equation}\label{BandEdges}
	\SpaceFunc_{\pm}\left(\Kb\right)\equiv\frac{2}{\Kb}\left(4+\Kb\pm\sqrt{2}\sqrt{8+\Kb}\right).
\end{equation}
When the values in~\cref{BandEdges} are substituted into~\cref{InverseFunction}, we obtain the radii~$\Risco\equiv r\left(\SpaceFunc_{\pm}\right)$
\begin{align}\label{ISCO}
	\Risco\left(\Rh,\Kb\right)\equiv & \frac{2\Rh\sqrt{4+\Kb\pm\sqrt{2}\sqrt{8+\Kb}}}{\sqrt{2}\sqrt{10+\Kb\pm2\sqrt{2}\sqrt{8+\Kb}}-2}
	\nonumber\\
	&\hspace{-70pt} \times\left[\frac{\sqrt{10+\Kb\pm2\sqrt{2}\sqrt{8+\Kb}}-\sqrt{2-\Kb}}{\sqrt{10+\Kb\pm2\sqrt{2}\sqrt{8+\Kb}}+\sqrt{2-\Kb}}\right]^{\frac{1}{\sqrt{4-2\Kb}}}.
\end{align}
Note that we can also substitute~\cref{HorizonFunction} into~\cref{ISCO} to obtain~$\Risco$ in terms of~$\Rg$ and~$\Kb$. Corresponding to~\cref{HorizonRange}, the ranges of the critical radii in~\cref{ISCO} are then
\begin{equation}
	\left(2\mp 1\right)\Rg<\Risco<\frac{\sqrt{3\pm\sqrt{5}}e^{\frac{2\sqrt{3\pm\sqrt{5}}-\sqrt{2}}{2\sqrt{3\pm\sqrt{5}}}}}{2\sqrt{3\pm\sqrt{5}}-\sqrt{2}}\Rg\equiv \zeta_{\pm}\Rg,\label{ISCORange}
\end{equation}
The upper bounds correspond to~$\zeta_{\pm}\Rg$ in~\cref{AllPlots}. From~\cref{ISCORange} we identify the physically interesting branch\footnote{We note that the second solution lies within the range~$\Rh<\Rp<\Rl$ for all~$\Kb$. The fact that this solution separates from the throat at all is a property of the E\AE{} solution that is not seen in the GR limit --- however in both theories~$\Rp$ corresponds to superluminal orbits.}~$r_-$ with the E\AE{} counterpart of the GR ISCO, which lies at~$3\Rg$ in the Schwarzschild solution of~\cref{SchwarzschildActual}.

\subsection{Formulation with a radial \ae ther component}\label{sec:GI}

The ansatz considered in the~\cref{sec:Jacobson} had an \ae ther field which was chosen to lie solely in the time direction. We want to show here how if this is generalised to allow a radial component, then of necessity this entails a cross-term in the metric, and thus a non-zero~$g_2$. We will then show, however, that in the new ansatz the equations can be expressed in terms of variables which are `gauge invariant' under the introduction of a radial component, meaning that the new solution is physically identical to the old one, and contains no additional physical information. This is consistent with the expectation that in a vacuum, the choice of `four-velocity' (which is basically what the~$A^\mu$ field represents here) is purely gauge. It is only when we have an additional physical velocity, such as the four-velocity of a fluid, to compare with, that this choice becomes a physical one. Prompted by this observation, we move on to consider a case in which a perfect fluid is indeed present, the four-velocity of which does provide a physical comparison, and we show that while we again have a gauge freedom left over, it still only represents a single degree of freedom, since (at least in our current setup) the directions of the~$A^\mu$ field and of the fluid velocity vector are obliged to coincide. This will be discussed in~\cref{sec:EAmatter}.

It is convenient to consider the components of the unit-length \ae{}ther vector in the local Lorentz basis at each point, for which we use the simple ansatz 
\begin{align}
[\tensor{A}{^a}]= [\cosh\alpha(r), \ \sinh\alpha(r),\ 0,\ 0].
\label{eq:generala}
\end{align}
Thus interpreted as a four-velocity,~$\tensor{A}{^a}$ corresponds to radial motion with a rapidity parameter~$\alpha(r)$ relative to the local Lorentz frame.

If we solve the time component of the \ae{}ther field equations for~$\alpha''$, and then substitute the result into the radial component, we obtain the simple relationship
\begin{align}
\lambda \left(g_1\sinh\alpha + g_2\cosh\alpha\right)=0.
\label{eqn:lam-gs-rel}
\end{align}
Thus the possibilities are that either~$\lambda=0$ or
\begin{align}
g_2=-g_1 \tanh\alpha.
\label{eqn:g2-expr}
\end{align}
By examining the remaining equations of motion, it turns out that~$\lambda=0$ leads to an analogue of the Reissner-Nordstr\"om solution, complete with the Schwarzschild geometry as its zero-charge limit. Therefore, we will assume~$\lambda\neq 0$ so that~\cref{eqn:g2-expr} applies. The~$t\theta$ component of the Einstein equations reads (up to factors which cannot vanish)
\begin{align}
\lambda r \cosh\alpha \sinh\alpha f_1 + g_2\left(g_1 f_1'+f_1 g_1'\right) =0,
\label{eqn:ein_ttheta}
\end{align}
from which, in conjunction with~\cref{eqn:g2-expr}, we deduce
\begin{align}
\lambda = \frac{g_1\left(g_1 f_1\right)'}{r f_1 \cosh^2\alpha }.
\label{eqn:lam_res}
\end{align}
We can now use~\cref{eqn:g2-expr} and~\cref{eqn:lam_res} in the \ae ther equations, which reduce to a single equation involving the first and second powers of~$f_1'$,~$\alpha'$ and~$g_1'$, and contain an overall factor of~$\Kb-2$. Assuming~$\Kb\neq 2$ (and thus avoiding the `extremal \ae ther limit in~\cref{KbRange}) we can use this equation to obtain an expression for the square of~$\alpha'$, which we then insert into the~$tt$ Einstein equation. This yields a simple equation which is linear in the first derivatives~$f_1'$,~$\alpha'$ and~$g_1'$
\begin{align}
    f_1 \cosh^3\alpha&-r g_1 f_1 g_1' \cosh\alpha-g_1^2 f_1 \cosh\alpha\nonumber\\
    &\hspace{10pt}+r g_1^2 f_1' \cosh\alpha+2 r g_1^2 f_1 \alpha'\sinh\alpha=0. \label{eq:f1_g1_alpha_prime}
\end{align}
Solving this equation for~$\alpha'$ and substituting the solution back into the Einstein and \ae ther equations then yields equations which have no~$\alpha$ derivatives, but which are still relatively complicated. However, there are some substitutions which simplify the equations significantly. We define two new variables~$X(r)$ and~$Y(r)$ according to
\begin{align}
X\equiv f_1 g_1, \quad Y\equiv\frac{1}{g_1}\cosh\alpha,
\label{eqn:X-Y-def}
\end{align}
and use these to substitute for~$f_1$ and~$\cosh\alpha$ in the equations. This yields a much simplified set of three equations
\begin{subequations}
\begin{align}
0=&2r Y' X + Y^3 X +r Y X'-YX,\label{eqn:alphad}\\
0=&\Kb(Y-1)^2(Y+1)^2 X^2\nonumber\\
	&-2 X'\left[4+\left(Y^2-1\right) \Kb\right] r X +\Kb r^2 X'^2,\label{eqn:ein-constraint}\\
0=&\Kb(Y-1)^2(Y+1)^2 X^2\nonumber\\
&+2 r\left[\left(X'' r+Y^2 X'+2 X'\right) \Kb-4 X'\right] X\nonumber\\
&-\Kb r^2 X'^2,\label{eqn:max-eqn}
\end{align}
\end{subequations}
respectively from the the~$\alpha'$, Einstein and \ae ther equations. There are other components to the Einstein and \ae ther equations, but they do not yield anything new beyond~\cref{eqn:ein-constraint,eqn:max-eqn}. Since there are no derivatives of~$Y$ in either of~\cref{eqn:ein-constraint,eqn:max-eqn}, it is possible to use these to get an equation involving~$X$ alone, and which is still second order, namely
\begin{align}
\Kb&\left(X'' r+3 X'\right)^2 X^2\nonumber\\
&+2 r\left[\left(X'' r+3 X'\right) \Kb-16 X'\right] X'^2 X\nonumber\\
&+\Kb r^2 X'^4 =0.
\label{eqn:X-eqn}
\end{align}
If the second order equation in~\cref{eqn:X-eqn} can be solved for~$X$, then one can show that using~\cref{eqn:alphad} it is possible to recover~$Y$ via
\begin{align}
	Y = \sqrt{r}\left[X\left(\displaystyle{\int \frac{\mathrm{d}r}{X}+ c}\right)\right]^{-\frac{1}{2}},
\end{align}
where~$c$ is a constant of integration, and we take the positive root since both~$g_1$ and~$\cosh\alpha$ have to be positive. Alternatively, one can use~\cref{eqn:alphad} to derive~$X$ in terms of~$Y$, for which the result is
\begin{align}
	X=\frac{Cr}{Y^2}\exp\left(-\int\frac{Y^2\mathrm{d}r}{r}\right),
	\label{eqn:X-expr}
\end{align}
where~$C$ is another constant. Substituting~\cref{eqn:X-expr} into~\cref{eqn:ein-constraint} then yields a first order equation in~$Y$ alone:
\begin{align}
-\Kb &Y^6+(2 \Kb-2) Y^4-2 \Kb Y^3 Y' r\nonumber\\
+&(2-\Kb) Y^2+2 Y' r(\Kb-2) Y-\Kb Y'^2 r^2=0 ,
\label{eqn:Y-alone-first-order}
\end{align}
which can be solved implicitly in terms of an integral in which~$Y(r)$ is the upper limit. One can check that the second order equation~\cref{eqn:max-eqn} is compatible with both these approaches, i.e.\ that we have a consistent set of equations.

Now, it will have been noticed that the definitions~\cref{eqn:X-Y-def} involve three variables as input, i.e.~$f_1$,~$g_1$ and~$\alpha$, but with \textit{two} variables,~$X$ and~$Y$ as output. However, there are no further independent equations in the system apart from those we have just written down for~$X$ and~$Y$. This means we cannot determine~$f_1$,~$g_1$ and~$\alpha$ individually --- there must be a gauge freedom between them meaning, for example, that we can choose~$\alpha$ as we wish, with the other two changing to accommodate this choice, but with~$X$ and~$Y$ remaining fixed.   

If we convert the expression for~$\lambda(r)$ given in equation~\cref{eqn:lam_res} to be in terms of our~$X$ and~$Y$ variables, we obtain
\begin{align}
\lambda=\frac{X'}{r X Y^2}.\label{eqn:lam-expr}
\end{align}
\cref{eqn:lam-expr} confirms that in general~$\lambda$ is `intrinsic', i.e.\ has the same physics attached to it regardless of the value of~$\alpha$. 

It is important to emphasize that although~\cref{eq:generala} has a radial component, the components~$\tensor{A}{^\mu} = \tensor{e}{_a^\mu} \tensor{A}{^a}$ in the coordinate basis possess only a time-like component after imposing~\cref{eqn:g2-expr}. This behavior stems from the equations of motion and the assumption that~$\tensor{A}{^\mu}$ depends solely on~$r$. For simplicity, one may consider this in Minkowski spacetime without any sources. The Maxwell equations read~$\left[\tensor{A}{^t}(r)' + r\tensor{A}{^t}(r)''\right]/r = 0$ and~$\tensor{A}{^r}(r)/r^2 = 0$, which immediately lead to~$\tensor{A}{^r} = 0$. This result can be generalised to electromagnetism in curved spacetime and E\AE{} theory, where the radial component still vanishes as derived above. This outcome is quite general and is independent of the condition~$\tensor{A}{^\mu} \tensor{A}{_\mu} = 1$. Thus, constraining the \ae{}ther field to be purely~$r$-dependent forces it to align exclusively with the time-like direction in the coordinate basis (and therefore be Lorentz-violating).

It is also worth highlighting how we can reach the expression of the equations in~$X$,~$Y$ form via a coordinate transformation from our setup involving an explicit~$\alpha$. Specifically we can do this via the following transformation of the time and space coordinates:
\begin{equation}
(t,r) \mapsto (t+f(r),r),
\label{eqn:coord-trans}
\end{equation}
where the function~$f(r)$ satisfies
\begin{equation}
f'(r) = \frac{f_1 \sinh(2\alpha)}{2g_1}.
\end{equation}
Integrating this with respect to~$r$ would require the assumption of a specific form for~$\alpha(r)$ of course, but the formula just given for the derivative of~$f$ is enough to show that under this transformation (and assuming the~$g_2 =- g_1 \tanh \alpha$ relationship enforced by the Maxwell equations), the metric loses its off-diagonal part, and becomes
\begin{align}
	\mathrm{d}s^2=\frac{1}{\left(f_1 \cosh\alpha\right)^2}\mathrm{d} t^2 &-\left(\frac{\cosh\alpha}{g_1} \right)^2\mathrm{d} r^2 \nonumber\\
 &\hspace{10pt}- r^2 \left(\mathrm{d}\theta^2 + \sin^2 \theta\mathrm{d} \phi^2  \right).
\end{align}
We recognise from the definitions in~\cref{eqn:X-Y-def} that 
\begin{align}
	\frac{1}{\left(f_1 \cosh\alpha\right)^2}=\frac{1}{\left(XY\right)^2},\quad\quad  \left(\frac{\cosh\alpha}{g_1} \right)^2 = Y^2
\end{align}
and so we have succeeded in reaching an intrinsic form of the metric. Everything in the earlier sections can now be accessed via the identifications
\begin{align}
\mathscr{R}=Y^2, \quad \mathscr{T}=\frac{1}{X^2 Y^2}.
\end{align}
Note the effect on the contravariant components of~$A$ due to the transformation~\cref{eqn:coord-trans} is to leave them invariant, i.e.~$\tensor{A}{^\mu}$ for~$\mu=0,1$ is given by~$1/\sqrt{\mathscr{T}}$ and 0, respectively, both before and after the transformation.

\section{The matter interior}\label{sec:matterin}
\subsection{The introduction of matter}\label{sec:EAmatter}
We carry on by looking at the case where a perfect fluid is introduced. We do this initially for the ansatz in ~\cref{eq:generala}, where the rapidity parameter~$\alpha(r)$ for the~$\tensor{A}{^a}$ unit vector, is allowed to vary. Additionally, we let the fluid velocity~$\tensor{v}{^a}$ in the local Lorentz basis have a radial component, with rapidity parameter~$\beta(r)$. We note that we have already shown in~\cref{sec:GI} that, in the case without matter, the introduction of non-zero~$\alpha$ is effectively just a gauge choice, with no physical consequences. Similarly, if we had no \ae ther component, then the introduction of~$\beta$ would not necessarily mean that the fluid had a genuine radial velocity. Indeed, with the accompanying introduction of a cross-term in the metric, this can again just be a choice of gauge. However, having both present means that the angle between them could in principle be measurable, so it is worth investigating this aspect, and having a suitably general setup from which to start.

The matter we introduce will be a perfect fluid with (Lorentzian) stress-energy tensor
\begin{align}
\tensor{T}{^{ab}} = (P + \rho) \tensor{v}{^a} \tensor{v}{^b} + P \tensor{\eta}{^{ab}},
\end{align}
where the four-velocity~$v$ is confined to the~$(t,r)$ plane and has rapidity parameter~$\beta(r)$. This~$\tensor{T}{^{ab}}$ multiplied by~$8 \pi$ is then added to the right hand side of the Einstein equations, but with the ansatz for other quantities remaining the same.

Here we provide a schematic account of how these equations can be treated. Firstly, as before, we solve the time component of the \ae ther equations for~$\alpha''$, and insert this into the radial component. Since the \ae ther equations are not directly affected by the presence of the matter, this yields~\cref{eqn:lam-gs-rel} as before, and since we are going to assume~$\lambda$ to be non-zero here, this once again leads to the expression for~$g_2$ in~\cref{eqn:g2-expr}. We can then use the~$tr$ Einstein equation to find an expression for~$\lambda$, and go through all the equations again substituting this for~$\lambda$ and~\cref{eqn:g2-expr} for~$g_2$. However, if we now take the new~$rr$ Einstein equation and solve for~$\alpha'^2$, and substitute this into the~$tt$ Einstein equation, we get the result (omitting factors which cannot vanish)
\begin{align}
\frac{\sinh\left[2(\alpha-\beta)\right](\rho+P)}{\sinh(2\alpha)}=0.
\end{align}
Thus, assuming~$\rho \neq -P$, if~$\alpha$ is non-zero\footnote{The conclusion~$\alpha=\beta$ remains true for the case~$\alpha=0$ despite the denominator being zero.} we must have~$\beta=\alpha$. We now take a similar route through to obtaining an equation which is linear in~$\alpha'$ as was already discussed following~\cref{eqn:lam_res} above. Again, we reinsert the solution for~$\alpha'$ in all the equations. At this stage, we find that despite the presence of the matter we can make exactly the same substitutions as in~\cref{eqn:X-Y-def}, which replace the three variables~$f_1$,~$g_1$ and~$\alpha$ with the two variables~$X$ and~$Y$. This means that the function~$\alpha(r)$ (or equivalently~$g_2$) is a pure gauge choice once again, since the physics is described by the reduced set~$X$ and~$Y$. We pick out two physical quantities of interest relative to the vacuum case, namely the pressure gradient~$\mathrm{d}P/\mathrm{d}r$ and the expression for the Lagrange multiplier~$\lambda$. We find
\begin{subequations}
\begin{align}
    P'&=\left[ \ln \left(XY \right) \right]' \left( \rho+P \right),\label{eq:XY_matter1} \\
    \lambda&=4\pi \left( \rho+P \right) + \frac{ \left[\ln \left(X\right)\right]'}{r Y^2}. \label{eq:XY_matter2}
\end{align}
\end{subequations}
The `vacuum' contribution to~$\lambda$ matches that given in~\cref{eqn:lam-expr}, whilst the matter contribution is a multiple of the pressure gradient, in the sense that both are proportional to~$\rho+P$.

\subsection{The governing equations}\label{sec:geqs}
In this section, we derive the governing equations that will be used in later discussions on Buchdahl's limit, and other numerical results. Specifically, we derive the Tolman–Oppenheimer–Volkoff (TOV) equation for the Einstein-\ae{}ther theory\footnote{See also~\cite{Reyes:2024oha} for the full \AE{}ST case.}, where we define an \ae{}ther energy that compactifies the final expression compared to~\cite{LEON2020168002}. Due to the gauge invariance discussed above, we are free to set~$\alpha=\beta=0$, simplifying the line element to that in~\cref{sec:Jacobson}. The resulting governing equations are
\begin{subequations}
\begin{align}
8\pi r^2\rho &= 1-\left(\frac{1}{\SpaceFunc}\right)-r\left(\frac{1}{\SpaceFunc}\right)'\nonumber\\
&\quad\quad+r \Kb\left[-\frac{r}{8}\left(\frac{\TimeFunc'}{\TimeFunc}\right)^2\left(\frac{1}{\SpaceFunc}\right) \right.\nonumber\\
&\quad\quad-\frac{r}{2}\left(\frac{1}{\SpaceFunc}\right) \left(\frac{\TimeFunc'}{\TimeFunc}\right)' -\left(\frac{1}{\SpaceFunc}\right) \left(\frac{\TimeFunc'}{\TimeFunc}\right) \nonumber\\
&\quad\quad\left. -\frac{r}{4}\left(\frac{\TimeFunc'}{\TimeFunc}\right)\left(\frac{1}{\SpaceFunc}\right)' \right], \label{eq:tteom} \\
8\pi r^2 P &= -1+\left(\frac{1}{\SpaceFunc}\right)+r\left(\frac{\TimeFunc'}{\TimeFunc}\right)\left(\frac{1}{\SpaceFunc}\right)\nonumber\\
&\quad\quad+\frac{\Kb}{8}r^2\left(\frac{\TimeFunc'}{\TimeFunc}\right)^2\left(\frac{1}{\SpaceFunc}\right), \label{eq:rreom}\\
8\pi r^2 P &=  \frac{1}{4}\left(\frac{\TimeFunc'}{\TimeFunc}\right)^2\left(\frac{1}{\SpaceFunc}\right) r^2-\frac{\Kb r^2}{8} \left(\frac{\TimeFunc'}{\TimeFunc}\right)^2\left(\frac{1}{\SpaceFunc}\right) \nonumber\\
&\quad\quad+\frac{r^2}{2}\left(\frac{1}{\SpaceFunc}\right)\left(\frac{\TimeFunc'}{\TimeFunc}\right)'+\frac{r}{2}\left(\frac{1}{\SpaceFunc}\right)'\nonumber\\
&\quad\quad+\left(\frac{\TimeFunc'}{\TimeFunc}\right)\left[\frac{r}{2}\left(\frac{1}{\SpaceFunc}\right) +\frac{r^2}{4}\left(\frac{1}{\SpaceFunc}\right)'\right], \label{eq:phiphieom} \\
P'&=-\frac{1}{2}\left(\frac{\TimeFunc'}{\TimeFunc}\right)(\rho+P). \label{eq:conserve_eom}
\end{align}
\end{subequations}
We can define the \ae{}ther energy as 
\begin{align}
\epsilon &\equiv \tensor{F}{^{0\mu}}\tensor{F}{^{0}_\mu}-\frac{1}{4}\tensor{g}{^{00}}\tensor{F}{^{\mu\nu}}\tensor{F}{_{\mu\nu}}= \frac{1}{32\pi}\left(\frac{\TimeFunc'}{\TimeFunc}\right)^2\left(\frac{1}{\SpaceFunc}\right).
\end{align}
Using the Schwarzschild parameterisation~$1/\SpaceFunc = 1-2m(r)/r$ and combining~\cref{eq:rreom} and~\cref{eq:phiphieom}, we obtain
\begin{align}
&r\left(\frac{\TimeFunc'}{\TimeFunc}\right)\left(\frac{1}{\SpaceFunc}\right) =m'+12\pi r^2 P - 8\pi r^2\epsilon 
\nonumber\\
	&\hspace{80pt}+2\pi \Kb r^2\epsilon-8\pi r^2\epsilon'\left(\frac{\TimeFunc'}{\TimeFunc}\right)^{-1}.\label{rphipsi}
\end{align}
Then, by combining~\cref{eq:tteom} and~\cref{rphipsi}, we can express the `mass' function as
\begin{align}
	m(r) = \int^r_0 \frac{8\pi\rho + 2\pi\Kb \left[ 6 P -(2-\Kb)\epsilon \right]}{2-\Kb} \bar{r}^2 \mathrm{d}\bar{r}.\label{apparent_mass}
\end{align}
One should note that to convert the `mass' here to the Schwarzschild mass, ensuring asymptotic flatness at infinity, one should substitute the radial metric into~\cref{InverseFunction} and~\cref{HorizonFunction}. From~\cref{eq:rreom} and~\cref{eq:conserve_eom}, one can derive the expression
\begin{align}
P' = -\frac{(P+\rho)(m + 4\pi r^3P -2\pi \Kb r^3\epsilon)}{r(r-2m)}.\label{eq:TOV_P_prime}
\end{align}
The final form of E\AE{} TOV equations are comprised of~\cref{apparent_mass} and~\cref{eq:TOV_P_prime}.

\subsection{Buchdahl's theorem}\label{sec:buchdahl}
\begin{figure}[ht!]
\includegraphics[width=8.6cm]{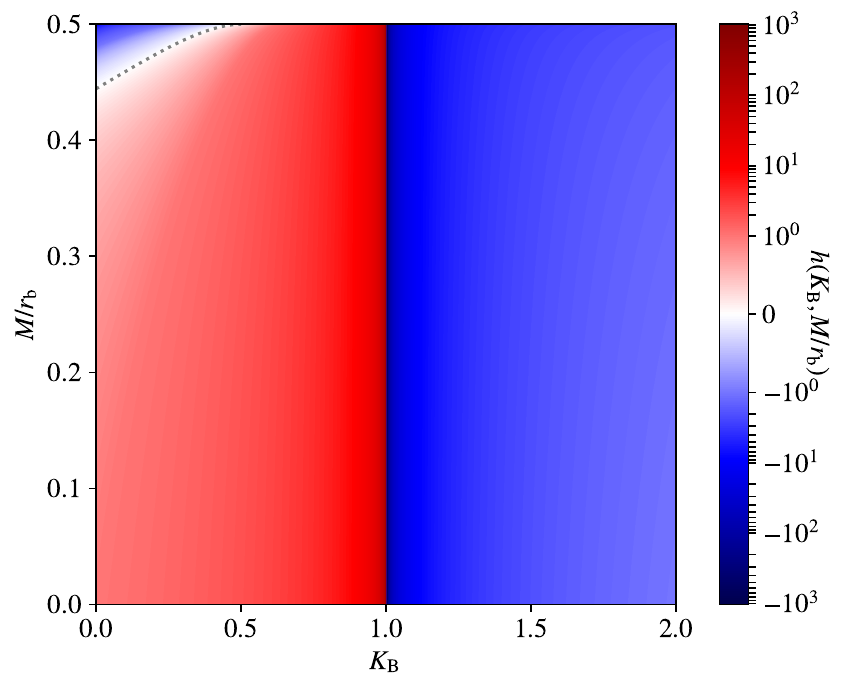}
	\caption{The change of~$h(\Kb,M/r_{\text{b}})$, which is the final inequality of Buchdahl theorem~\cref{eq:final_ineq} with~$0$ being the limit, with different~$\Kb$ and~$M/r_{\text{b}}$. The dotted line is the analytical solution from~\cref{buchdahl}. It can be seen that there is a white region near~$h(\Kb,M/r_{\text{b}})=0$ between~$0<\Kb\leq 1/2$, which corresponds to~\cref{buchdahl}. For the remaining region, there is no additional bound from~\cref{eq:final_ineq}.}
\label{fig:buchdahl_map}
\end{figure}

We are now in a position to derive Buchdahl's theorem, which describes the limit of compactness for static, spherically symmetric interiors. In GR, this limit is~$M_{\text{Sz}}/r_{\text{b}} \leq 4/9$, where~$r_{\text{b}}$ is the star radius (the radius beyond which lies the Eling--Jacobson vacuum). To proceed, we assume the following conditions within the star:~$\mathrm{d}(m/r^3)/\mathrm{d}r \leq 0$,~$m \geq 0$,~$\TimeFunc(r=0)\geq 0$, and~$\mathrm{d}\TimeFunc/\mathrm{d}r \geq 0$, which qualify a stable spherical star as outlined in~\cite{PhysRevD.92.064002, Wald:1984rg}. It should be noted that we do not assume any equation of state while deriving the theory as~\cite{Coley:2019aa} does. To make use of the condition of monotonically decreasing effective density, we should rewrite the equations of motion so that~$m/r^3$ shows up in the equation. Once this is done, one should find an expression involving~$\TimeFunc(r)$ at centre and boundary so as to constrain the expression between~$0$ and~$1$, which can be exploited to solve for the final Buchdahl's limit. By combining~\cref{eq:rreom} and~\cref{eq:phiphieom} to eliminate~$P$, and after some algebraic manipulations, we obtain
\begin{align}
\frac{\mathrm{d}}{\mathrm{d}r} \left[ \frac{1}{r\sqrt{\SpaceFunc\TimeFunc^{\Kb}}} \frac{\mathrm{d}\TimeFunc^{1/2}}{\mathrm{d}r} \right] = \sqrt{\TimeFunc^{(1-\Kb)}\SpaceFunc} \frac{\mathrm{d}}{\mathrm{d}r}\left[ \frac{m(r)}{r^3} \right].
\label{G11G22}
\end{align}
The condition~$\mathrm{d}(m/r^3)/\mathrm{d}r \leq 0$ implies that the left-hand side of~\cref{G11G22} is less than or equal to zero. Integrating this from~$r$ to~$r_{\text{b}}$ yields
\begin{align}
 \frac{1}{r\sqrt{\SpaceFunc\TimeFunc^{\Kb}}} \frac{\mathrm{d}\TimeFunc^{1/2}}{\mathrm{d}r} &\geq \frac{1}{2r_{\text{b}}\sqrt{\SpaceFunc_\text{b}\TimeFunc_\text{b}^{\Kb+1}}} \frac{\mathrm{d}\TimeFunc(r_{\text{b}})}{\mathrm{d}r}  \nonumber \\
&= \frac{1}{2r_{\text{b}}\sqrt{\SpaceFunc_\text{b}\TimeFunc_\text{b}^{(\Kb-1)}}}\left(\frac{\TimeFunc'}{\TimeFunc}\right)_\text{b}, 
\label{1stint}
\end{align}
where the subscript `b' indicates that the quantity is evaluated at the star surface. To address the LHS of~\cref{1stint}, we need
\begin{align}
 \int^{r_{\text{b}}}_0 &\left(\sqrt{\TimeFunc}\right)^{-\Kb} \frac{\mathrm{d}\TimeFunc^{1/2}}{\mathrm{d}r} \nonumber\\
 &= \frac{1}{1-\Kb}\left( \sqrt{\TimeFunc_\text{b}}^{(1-\Kb)} - \sqrt{\TimeFunc_\text{c}}^{(1-\Kb)}\right),
 \end{align}
where the subscript `c' denotes that the quantity is evaluated at~$r=0$. This expression is only valid if~$\Kb\neq 1$\footnote{The calculation can be repeated for the~$\Kb=1$ case. Still, we find that no further constraint will be put in this particular case.}. Multiplying~\cref{1stint} by~$r\sqrt{\SpaceFunc}$ and integrating from~$0$ to~$r_{\text{b}}$, we obtain
\begin{align}
&\frac{1}{1-\Kb}\left(\sqrt{\frac{\TimeFunc_\text{c}}{\TimeFunc_\text{b}}}\right)^{(1-\Kb)} \nonumber\\
	&\leq \frac{1}{1-\Kb} - \frac{1}{2r_{\text{b}}\sqrt{\SpaceFunc_\text{b}}}\left(\frac{\TimeFunc'}{\TimeFunc}\right)_\text{b} \int_0^{r_{\text{b}}} \frac{r\mathrm{d}r}{\sqrt{ 1- 2m(r)/r}}  \nonumber \\
	&\leq  \frac{1}{1-\Kb} - \frac{1}{2r_{\text{b}}\sqrt{\SpaceFunc_\text{b}}}\left(\frac{\TimeFunc'}{\TimeFunc}\right)_\text{b} \int_0^{r_{\text{b}}} \frac{r\mathrm{d}r}{\sqrt{ 1- 2Mr^2/r_{\text{b}}^3}} \nonumber \\
&=  \frac{1}{1-\Kb} - \frac{1}{2r_{\text{b}}}\sqrt{1-\frac{2M}{r_{\text{b}}}}\left(\frac{\TimeFunc'}{\TimeFunc}\right)_\text{b}  \nonumber \\
	&\hspace{90pt}\times\frac{r_{\text{b}}^3}{2M}\left(1-\sqrt{1-\frac{2M}{r_{\text{b}}}}\right).
\label{2ndint}
\end{align}
The second inequality arises from the assumption on the mass, namely~$m \geq Mr^3/r_{\text{b}}^3$, with~$M\equiv m(r_{\text{b}})$. In the final equality, we insert~$1/\SpaceFunc_\text{b} = 1-2M/r_{\text{b}}$, which itself sets the constraint that~$M/r_{\text{b}}\leq1/2$.  Moreover,~$(\TimeFunc'/\TimeFunc)_{\text{b}}$ can be determined from~\cref{eq:rreom} with vanishing fluid pressure~$P=0$ at the boundary
\begin{align}
&\left(\frac{\TimeFunc'}{\TimeFunc}\right)_\text{b} 
	= 4\left[ \frac{\sqrt{1-\frac{(2-\Kb)M}{r_{\text{b}}}}-\sqrt{1-\frac{2M}{r_{\text{b}}}}}{\Kb r_{\text{b}}\sqrt{1-\frac{2M}{r_{\text{b}}}}}\right]. 
\label{phi_b}
\end{align}
A thorough discussion of why~$P_\text{b}=0$ is presented in~\cref{app:P_fl}. With these observations, and with~\cref{phi_b}, it is possible to rewrite~\cref{2ndint} as
\begin{align}
    &\frac{1}{1-\Kb}\left(\sqrt{\frac{\TimeFunc_\text{c}}{\TimeFunc_\text{b}}}\right)^{(1-\Kb)} \nonumber\\
    &\leq\frac{1}{1-\Kb}-\frac{r_{\text{b}}}{M\Kb}\left(1-\sqrt{1-\frac{2M}{r_{\text{b}}}}\right)\nonumber\\
    &\quad\quad\times\left(\sqrt{1-\frac{\left(2-\Kb\right)M}{r_{\text{b}}}}-\sqrt{1-\frac{2M}{r_{\text{b}}}}\right) \nonumber \\
    &\equiv h(\Kb,M/r_{\text{b}}),\label{eq:final_ineq}
\end{align}
where the function~$h(\Kb,M/r_{\text{b}})$ is introduced as a useful definition. For~$0\leq\Kb<1$, the LHS of~\cref{eq:final_ineq} is larger than or equal to~$0$; therefore, we can solve the inequality that~$h(\Kb,M/r_{\text{b}})$ is greater than or equal to~$0$ and thereby obtain Buchdahl's theorem. As this inequality depends only on~$M/r_{\text{b}}$, it can be solved to yield the final bound on~$M/r_{\text{b}}$, namely
\begin{align}
\frac{M}{r_{\text{b}}} \leq \frac{4 \left(1-\Kb\right)}{\left(3-2 \Kb\right)^2}.
\label{buchdahl}
\end{align}
One should note that this expression is valid only for~$0 \leq \Kb \leq 1/2$. For~$1/2 < \Kb < 2$, no further constraint is imposed, and the only condition in this regime is~$0 < M/r_{\text{b}} < 1/2$. This is illustrated numerically in~\cref{fig:buchdahl_map}, which plots the function~$h(\Kb, M/r_{\text{b}})$ over the entire allowed parameter space. The white region, where~$h(\Kb, M/r_{\text{b}}) = 0$, corresponds to Buchdahl's limit in~\cref{buchdahl}. For~$1/2 < \Kb < 1$,~$h(\Kb, M/r_{\text{b}})$ is positive throughout the entire range, indicating no constraint other than~$M/r_{\text{b}}\leq1/2$ (which follows from definition). Similarly, since the left-hand side of~\cref{eq:final_ineq} becomes negative and unbounded for~$\Kb > 1$, no additional constraint applies in this part of the parameter space either. Analytically, by inserting~\cref{buchdahl} into the expression for~$h(\Kb, M/r_{\text{b}})$, one finds that~$h(\Kb, M/r_{\text{b}}) \propto -1 + 2\Kb + |1 - 2\Kb|$ for~$0 < \Kb < 1$, implying that solutions only exist for~$0 < \Kb < 1/2$. A physical explanation would be, for~$\Kb>1/2$, the throat where~$\SpaceFunc \mapsto \infty$ sits inside the object, which will be further discussed in~\cref{app:unphysical}. We note once again that the~$M$ is \textit{not} the Schwarzschild mass~$M_{\text{Sz}}\equiv r_{\text{Sz}}/2$, but it can be converted to the Schwarzschild mass using~\cref{InverseFunction} and~\cref{HorizonFunction}. The resulting constraint is illustrated in~\cref{fig:buchdahl}. One can see that even if there is no constraint on~$M/r_{\text{b}}$, the bound on~$M_{\text{Sz}}/r_{\text{b}}$ is still lowered due to the definition of the Schwarzschild mass.

\subsection{The saturated mass profile}\label{sec:const_m_r}
In this section, a saturated case of the Buchdahl's theorem is discussed, in which~$m(r)/r^3$ is constant. Under this assumption, we define~$\SpaceFunc\equiv 1/(1-k r^2)$, where~$k$ is a constant. With these ans\"atze there exists an analytical solution
\begin{subequations}
\begin{align}
    \frac{\TimeFunc'}{\TimeFunc}&=\frac{2 k r}{k \left(r^2+2 c_1 \sqrt{1-k r^2}\right)+\Kb-1-k \Kb r^2},\\
    P &= k \left(k r^2-1\right) \bigg[8 k^2 c_1{}^2+\Kb^2 \left(2-2 k r^2\right) \nonumber\\
    &\quad\quad+\Kb \left(7 k r^2+8 k c_1 \sqrt{1-k r^2}-8\right) \nonumber\\
    &\quad\quad-2 k \left(3 r^2+8 c_1 \sqrt{1-k r^2}\right)+6\bigg] \nonumber\\
	&\quad\quad\times \frac{1}{16\pi }\bigg[\Kb \left(k r^2-1\right)\nonumber\\
	&\quad\quad-k \left(r^2+2 c_1 \sqrt{1-k r^2}\right)+1\bigg]^{-2}. \label{eq:extreme_case_analytical_sol}
\end{align}
\end{subequations}
We force the pressure to vanish at the boundary~$r_{\text{b}}$, and solve for~$c_1$ to obtain
\begin{align}
	c_1 = \frac{1}{2k}\sqrt{1+\frac{(\Kb-2)M}{r_{\text{b}}}}-\frac{\Kb-2}{2k}\sqrt{1-\frac{2M}{r_{\text{b}}}}.
\end{align}
To solve the extreme case, one should calculate~$M/r_{\text{b}}$ when the denominator of pressure vanishes. This leads to
\begin{align}
    \frac{M}{r_{\text{b}}}=\frac{4 (1-\Kb)}{(3-2 \Kb)^2}. \label{eq:m_rb_const}
\end{align}
This result is only valid if~$\Kb\le 1/2$ due to the same reasoning as presented in~\cref{sec:buchdahl,app:unphysical}; that is, the throat is inside the object. Accordingly, there is no extreme case for the alternative region. Furthermore,~\cref{eq:m_rb_const} matches the Buchdahl theorem in~\cref{buchdahl}, and is thus consistent with our derivation in~\cref{sec:buchdahl}. On the other hand, one should note that the solution presented here is from the physical branch, which satisfies the strong energy condition. A wrong branch choice would render the solution unable to match the asymptotically flat spacetime as discussed in~\cref{app:unphysical}.

\section{Numerical results}\label{sec:GC}
In this section, we numerically examine E\AE{} theory across different scales. We begin by considering a uniform physical density as a toy model of a neutron star with which to illustrate Buchdahl's theorem. Subsequently, we present an analysis of the radial acceleration relation (RAR),  comparing the results with the Newtonian relation. To address these scenarios, we assume an exponentially decreasing density profile given by
\begin{align}
    \rho = \rho_0 e^{-u_0 r},
\end{align}
where~$u_0$ is a constant that characterizes the length scale. In this analysis, we set~$\alpha = 0$ for simplicity. The evolution equations remain the same as in~\cref{eq:tteom,eq:rreom,eq:conserve_eom}. To solve for the pressure and metric functions, we perform a series expansion around~$r=0$. The resulting solutions are
\begin{subequations}
\begin{align}
	\TimeFunc &= \frac{8\pi\left(3P_0+\rho_0\right)r^2}{3\left(2-\Kb\right)} - \frac{4\pi\rho_0 u_0 r^3}{3\left(2-\Kb\right)} + O\left(r^4\right), \label{eq:series_exp_matter1}\\
	\SpaceFunc &= 1 + \frac{8\pi\left(3\Kb P_0 + 2\rho_0\right)r^2}{3\left(2-\Kb\right)} 
\nonumber\\
	&\hspace{30pt}	- \frac{4\pi\rho_0 u_0 r^3}{2-\Kb} + O\left(r^4\right), \label{eq:series_exp_matter2}\\
	P &= P_0 - \frac{4\pi\left(3P_0^2 + 4P_0 \rho_0 + \rho_0^2\right)r^2}{3\left(2-\Kb\right)} \nonumber\\
&\hspace{30pt} + \frac{2\pi\left(15P_0 \rho_0 u_0 + 7\rho_0^2 u_0\right)r^3}{9\left(2-\Kb\right)} + O\left(r^4\right),
\label{eq:series_exp_matter3}
\end{align}
\end{subequations}
where~$P_0$ is the pressure at the origin. This initial pressure~$P_0$ must be chosen carefully to ensure that the pressure vanishes at infinity. Further discussion on this point is provided in~\cref{sec:NuBuchdalh}.

\subsection{Uniform physical density}\label{sec:NuBuchdalh}
\begin{figure}[ht!]
\includegraphics[width=8.6cm]{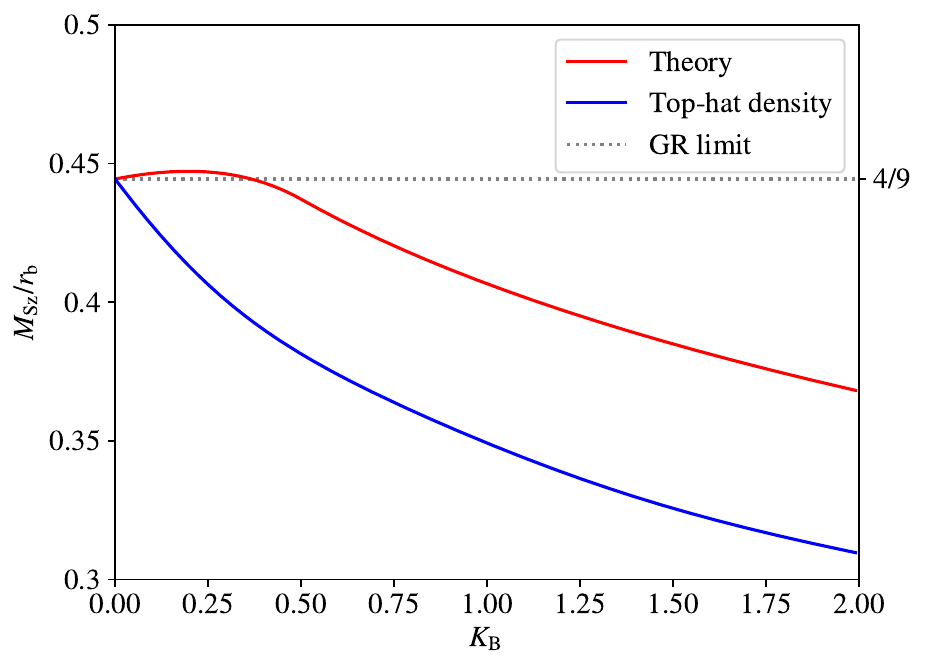}
	\caption{The variation of~$M_{\text{Sz}}/r_{\text{b}}$ with different values of~$\Kb$. The red curve represents the theoretical prediction derived from~\cref{buchdahl}, while the blue curve corresponds to the result from numerical integration, where a uniform physical density is assumed. The horizontal grey line marks the GR limit of~$4/9$ for comparison. The theoretical prediction indicates that the bound slightly exceeds the GR limit for small~$\Kb$, while it falls below the GR limit as~$\Kb$ increases. In the case of uniform density, the bound is consistently stricter than the GR limit.}
\label{fig:buchdahl}
\end{figure}

	In this section, we discuss a top-hat density profile with~$u_0 = 0$. This profile represents the extreme case in ordinary general relativity (GR) and is often used in GR textbooks as a heuristic model for neutron stars. It could be difficult to compute the extreme profile in E\AE{} theory since the integration is from~$r=0$, where the pressure~$P_0$ is required to be infinite by construction. However, due to the property of saturation, as long as~$P_0$ is chosen to be large enough, the radius turns out to vary only mildly with respect to different initial pressure choices. We select~$P_0$ to be~$10^3$ times larger than the density, which is sufficiently large for our case while also enhancing the computational efficiency and accuracy (larger~$P_0$ requires finer integration steps around the centre). For example, if one chooses an even larger~$P_0$ being~$10^5$ times of the density, the surface radius only varies at the order of~$0.1\%$. It should be noted once again that the star surface is defined to be the radius where the pressure vanishes numerically. The blue line in~\cref{fig:buchdahl} illustrates the limit for this scenario. It is evident that the saturation bound is lower for larger~$\Kb$ values. This suggests a trend opposite to the na\"ive expectation that the `weight' of the additional \ae{}ther field enhances the limit, as happens in other modified gravity theories which augment the Einstein--Hilbert term with additional operators (see e.g.~\cite{PhysRevD.92.064002}). On the other hand, this case is also an illustration of the validity of the Buchdahl theorem derived in~\cref{sec:buchdahl}. One can confirm that the uniform density case matches all the condition required for deriving the theorem. As in~\cref{fig:buchdahl}, the uniform density case lies below the Buchdahl limit. Accordingly, uniform physical density is \textit{not} the extreme case of E\AE{} theory.

\subsection{Radial Acceleration Relation}\label{sec:NuGC}
\begin{figure}[ht!]
\includegraphics[width=8.6cm]{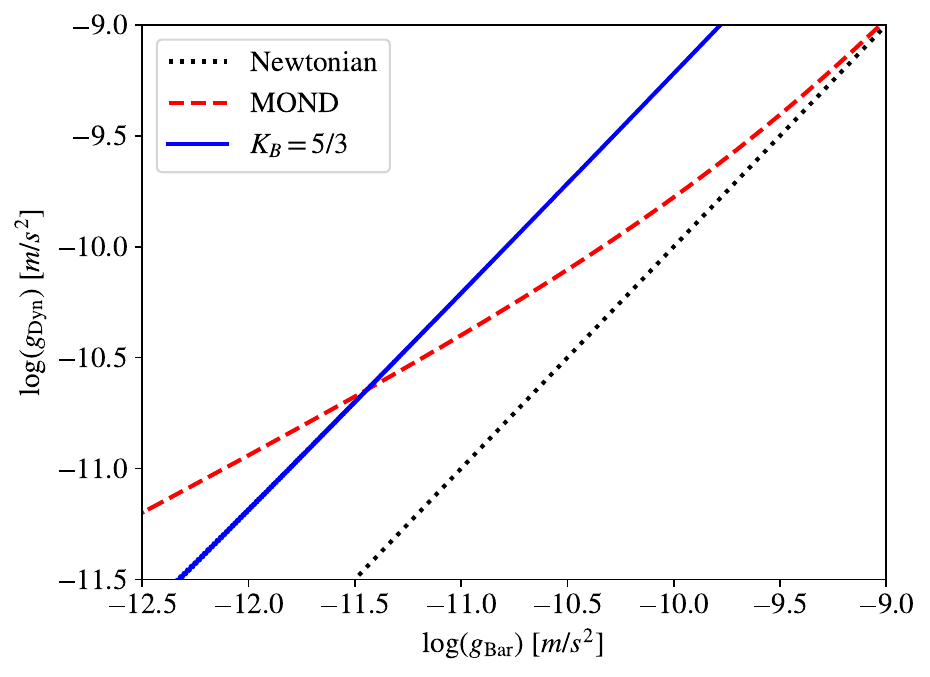}
	\caption{Radial acceleration relation (RAR), where~$g_{\text{Dyn}}$ and~$g_{\text{Bar}}$ are dynamical acceleration and Newtonian acceleration with baryons only, respectively. The blue line shows a numerical evaluation of the E\AE{} case with~$\Kb=5/3$, while the black dotted line shows the Newtonian case which runs straight to the origin with a slope near unity. The red dashed line shows a prediction from MOND --- see e.g.\ ~\cite{Famaey:2011kh}. The E\AE{} profile is roughly six times larger than the Newtonian one and parallel to it, which verifies the scaling relation~\cref{eq:scaling} for this particular~$\Kb$.}
\label{fig:acc}
\end{figure}

In this section we will use a combination of numerical results and our analytic series expansion, to examine the radial acceleration relation (RAR) for an extended object such as a galaxy or cluster of galaxies. Here one compares the dynamical acceleration as inferred from an indicator such as the azimuthal velocity, with the Newtonian acceleration expected from the enclosed baryonic mass. In regions of low intrinsic acceleration then MOND-type theories predict a higher dynamical acceleration than would be expected on a Newtonian basis, leading to curves such as the red-dashed one in~\cref{fig:acc} (see e.g.\ the review~\cite{Famaey:2011kh}). The range of `baryonic' versus dynamical accelerations in this figure is appropriate to that encountered in galaxies and clusters of galaxies, and our calculations are directed to these scales. Of course an exponential density profile, as we have used for the series expansions, is not very realistic, but has the advantage of being simple analytically, and allowing interpolation between on the one hand, the case of constant density, as we used in the neutron star toy model, and on the other, an object that does not have an explicit cutoff, but does have a convergent mass, as appropriate for galaxies and clusters.

Here, we start by inserting the series expansion~\cref{eq:series_exp_matter1} into the expression for azimuthal velocity squared given in~\cref{VelocityFunction}, yielding \begin{align}
    v^2 = \frac{8 \pi r^2 \left(3 P_0 + \rho_0\right)}{3\left(2-\Kb \right)} - \frac{2 \pi \rho_0 u_0 r^3}{2-\Kb} + O\left(r^4\right).\label{eq:v2_series}
\end{align}
The general relativistic (GR) case can be recovered by setting~$\Kb = 0$. Considering the function~$v^2/r$, which gives the inferred dynamical acceleration, it is evident that the first two terms in the GR and E\AE{} cases will yield the following scaling relation:
\begin{align}
    g_{E\textnormal{\AE}} \approx \frac{2}{2 - \Kb} g_N, \label{eq:scaling}
\end{align}
Considering higher-order terms, the~$r^4$ terms differ slightly, with the discrepancy proportional to~$\rho_0 P_0$, which is expected to be small compared to terms proportional to~$\rho_0$ and the same applies to the~$r^5$ term. Therefore, we predict that, in the limit of low, but not necessarily zero pressures, the velocities for the two cases will differ by a constant ratio over a range of~$r$s. Indeed, our numerical findings support this, showing that this ratio is maintained across the entire range of~$r$ of interest. This result is shown in the blue and dotted black curves of~\cref{fig:acc}, which are for the E\AE{} 
and Newtonian accelerations respectively, for the specific choice~$\Kb=5/3$, and where the (small) effects of pressure are included. Here our scaling relation would predict a ratio of around 6 in accelerations, as is seen in the diagram. We note that a relation of this kind is in agreement with the analysis carried out in the paper~\cite{PhysRevD.70.123525} by Carroll and Lim, via a different method, in which (assuming zero pressure), a Newtonian limit of the~$tt$ component of the Einstein equations was taken and they found that the effective value of Newton's constant~$G$ in the resulting Poisson equation changed by a constant ratio.

One should note that E\AE{} alone cannot address astrophysical observations of galaxies and clusters. On the galactic scale, due to the parallelism between the Newtonian and E\AE{} relations, one cannot produce flat or rising galaxy rotation curves. On the scale of galaxy clusters, the constant ratio in accelerations bears some relation to what is observed (see e.g.\ ~\cite{refId0}), but there is no reason within the theory for the particular value of~$\Kb$ which would be necessary to make this work.  Moreover, since~\cref{eq:scaling} renormalises the Newtonian constant on all scales, it may in principle lead to no effective change. A more realistic theory could possibly incorporate an effective~$\Kb$, which changes with scale, but this would be a matter for future work. 

\section{Conclusions}\label{sec:conclusion}
The main results of this paper are as follows:
\begin{itemize}
	\item We discuss the vacuum and matter solution of E\AE{} theory. In~\cref{Killing,TIntegral,InverseFunction,HorizonFunction} we have presented the exact non-rotating vacuum solution for the minimal E\AE{} theory of~\cref{EAeLagrangian}. As shown in~\cref{AllPlots}, our solution smoothly connects to the Schwarzschild BH of GR. As the E\AE{} parameter~$\Kb$ increases from the GR limit to the `extremal \ae ther' limit, the Newtonian limit of the E\AE{} BH remains unchanged, whilst the central nonlinear structures --- the throat in~\cref{HorizonFunction}, the ISCO in~\cref{ISCO} and the photon ring in~\cref{Luminal} --- become \textit{magnified}. Such exact formulae are an attractive route for observational constraints on modified gravity~\cite{EventHorizonTelescope:2021dqv}, in light of recent horizon-scale images of the presumed BHs M87*~\cite{EventHorizonTelescope:2019dse} and Sgr A*~\cite{EventHorizonTelescope:2022wkp}.
    \item In~\cref{sec:GI,sec:EAmatter}, a gauge freedom in spherically symmetric E\AE{} theory was established. By selecting a new set of parameters in~\cref{eqn:X-Y-def}, which includes three metric variables, the evolution equations can be written by these parameters exclusively as in~\cref{eqn:alphad,eqn:ein-constraint,eqn:max-eqn,eq:XY_matter1,eq:XY_matter2}. Since there are only two physical quantities, but three metric variables, there is \textit{one} residual gauge freedom. This freedom allows us to impose that the mixed metric component vanishes, thereby simplifying the calculations. 
	\item The analysis with the presence of perfect fluid begins with Buchdahl's theorem, leading to~\cref{eq:final_ineq}. Further investigation of this equation yields~\cref{fig:buchdahl_map}, where it can be observed that, for~$\Kb \leq 1/2$, a new constraint~\cref{buchdahl} is found. For other values of the coupling constant, no additional constraint other than~$M/r_{\text{b}}<1/2$ is imposed by~\cref{eq:final_ineq}. One should note that~$M$ is not the Schwarzschild mass, a conversion between these two definitions is required. After performing the conversion, as shown in~\cref{fig:buchdahl}, it becomes evident that Buchdahl's bound is \textit{lowered}. Only in the low-$\Kb$ region does the bound exceed the GR limit of~$4/9$.
	\item To investigate Buchdahl's theorem further in a more physical scenario, we consider a toy model of a neutron star with uniform physical density. The numerical results for this case are shown in~\cref{fig:buchdahl}. It is evident that the curve lies entirely below the GR limit for all allowed values of~$\Kb$, which may be opposite to the expectation that Buchdahl's bound would be relaxed as in other modified gravity theory~\cite{PhysRevD.92.064002}.
	\item We find a scaling relation of the RAR in~\cref{eq:scaling}, where the E\AE{} profile runs parallel to the Newtonian one. From the analytical series expansion~\cref{eq:v2_series}, one arrives at this scaling relation. Numerically, the parallelism is shown in~\cref{fig:acc}.  This feature may be of interest in relation to the actual RARs of galaxy clusters as described in~\cite{Wu:2014eva,Lelli:2016cui,refId0,Tian_2020,Eckert_2022}. However, this requires that~$\Kb$ varies across different scales, and further research should be done to understand this possibility.
\end{itemize}

\begin{acknowledgements}
We are grateful for useful discussions with Tobias Mistele, Yang Lirui and Tom Złośnik.

    Y-HH is supported by the doctoral scholarship from Taiwan Ministry of Education. WB is grateful for the support of Girton College, Cambridge, Marie Skłodowska-Curie Actions and the Institute of Physics of the Czech Academy of Sciences. AD was supported by the European Regional Development Fund and the Czech Ministry of Education, Youth and Sports: Project MSCA Fellowship CZ FZU I – CZ.02.01.01/00/22 010/0002906.
\end{acknowledgements}
\bibliographystyle{apsrev4-1}
\bibliography{Main,Manuscript,ManualRefs}

\appendix
\section{Matching the interior and exterior}\label{sec:appendix}
In this appendix, we will
\begin{enumerate}[label=(\alph*)]
  \item Consider the requirement on fluid pressure for matching at the object boundary;
  \item Discuss the exterior vacuum solutions that can be matched to, and in particular show how one branch is given by a form of inverse transform of the other;
  \item Briefly discuss an analytic solution for a specific~$\Kb$, which illustrates that for unphysical solutions, matching with a Schwarzschild-like exterior solution may not be possible.
\end{enumerate}

\subsection{Matching conditions at the boundary}\label{app:P_fl}

We want to discuss matching at the boundary,~$r=r_\text{b}$, and will do this by considering the behaviour of equations~\cref{eq:tteom} through~\cref{eq:conserve_eom} either side of the boundary. Let us introduce a new notation
\begin{equation}
\frac{\TimeFunc'}{\TimeFunc}=\phi, \quad 
\SpaceFunc = \frac{1}{\psi}
\end{equation}
In terms of these variables, the r.h.\ sides of~\cref{eq:tteom} through~\cref{eq:conserve_eom}  can be written more succinctly. We see in particular that~\cref{eq:tteom} and~\cref{eq:phiphieom} can be inverted to give the first derivatives of~$\phi$ and~$\psi$ in terms of~$\phi$,~$\psi$,~$\rho$ and~$P$. Then~\cref{eq:rreom}  can be used to get~$P$ in terms of~$\phi$ and~$\psi$ and so we now have expressions for~$\phi'$ and~$\psi'$ in terms of just~$\phi$,~$\psi$ and~$\rho$. In general, and in particular in the uniform case,~$\rho$ will have step at the boundary. This feeds through to show us that~$\phi$ and~$\psi$ must be continuous at the boundary, but with a step in their first derivatives.

Next, looking at~\cref{eq:rreom} and~\cref{eq:conserve_eom}, which give~$P$ and its derivative, we see that~$P$ has the same behaviour as~$\phi$ and~$\psi$, i.e.~$P$ must also be continuous at the boundary, but its first derivative will in general have a step. There is no fluid outside the boundary, hence continuity tells as that as~$P$ approaches the boundary from inside, then~$P\mapsto 0$ there. We note in particular that it is the \textit{fluid} pressure that must be zero at the boundary, not some assumed \textit{ effective} pressure equal to the sum of the fluid and `\ae ther' pressures. 

This result seems to be in contrast to the setup assumed in reference~\cite{PhysRevD.92.064002}, which while not working in the Einstein \AE ther theory is nevertheless dealing with a case,~$f(R)$ gravity, where there are contributions to the stress-energy tensor from both fluid pressure, and another pressure which they say is `sourced by the scalar curvature and its derivatives'. In their Section III.B (`Matching conditions'), they say that `matching the second fundamental form dictates that
the total radial pressure at the surface of the star must
vanish', where this pressure includes both contributions, as one can see from their equation (10). Here we believe that the equations of motion must be the primary way for the matching conditions to be established, and that this leads to the fluid~$P$ at the boundary being required to vanish.

\subsection{Branches of vacuum solutions}\label{app:branch}
In~\cref{sec:Jacobson}, we derived the vacuum solution. However, it is important to note that a specific branch was selected during the discussion. To identify the origin of these branches, we should derive a vacuum equation involving only~$\SpaceFunc$ by setting the fluid density~$\rho$ and pressure~$P$ to zero in~\cref{eq:tteom,eq:rreom,eq:phiphieom}. By combining~\cref{eq:tteom} and~\cref{eq:phiphieom},~$(\TimeFunc'/\TimeFunc)'$ can be eliminated, yielding an expression for~$(\TimeFunc'/\TimeFunc)$ in terms of~$\SpaceFunc$ and~$r$. By substituting this expression into~\cref{eq:rreom}, and inverting~$\SpaceFunc'(r)$ to~$\mathrm{d}r(\SpaceFunc)/\mathrm{d}\SpaceFunc$, we obtain
\begin{align}
    &\Kb r(\SpaceFunc)^2 + 8 \SpaceFunc r(\SpaceFunc) r'(\SpaceFunc) - 4 \Kb \SpaceFunc r(\SpaceFunc) r'(\SpaceFunc) \nonumber\\
    &\quad+ 4 \Kb \SpaceFunc^2 r(\SpaceFunc) r'(\SpaceFunc) - 8 \SpaceFunc^2 r'(\SpaceFunc)^2 \nonumber\\
    &\quad+ 4 \Kb \SpaceFunc^2 r'(\SpaceFunc)^2 + 8 \SpaceFunc^3 r'(\SpaceFunc)^2 \nonumber\\
    &\quad- 8 \Kb \SpaceFunc^3 r'(\SpaceFunc)^2 + 4 \Kb \SpaceFunc^4 r'(\SpaceFunc)^2 = 0,
    \label{eq:rR_equation}
\end{align}
which is an equation involving only the variable~$\SpaceFunc$ with the radial derivative~$r'(\SpaceFunc)\equiv\mathrm{d}r(\SpaceFunc)/\mathrm{d}\SpaceFunc.$ It is important to note that~\cref{eq:rR_equation} is only valid for~$\Kb \neq 1$. This equation exhibits symmetry under the transformation
\begin{align}
    r(\SpaceFunc) \mapsto \frac{C}{r(\SpaceFunc)} \, \frac{\SpaceFunc}{\SpaceFunc-1},
    \label{eqn:branch-relation}
\end{align}
where~$C$ is an arbitrary constant. This symmetry can be further explained by solving~\cref{eq:rR_equation} for~$r'(\SpaceFunc)$, yielding two solutions:
\begin{subequations}
\begin{align}
	r'(\SpaceFunc) =& \frac{\Kb r(\SpaceFunc)/2\SpaceFunc}{\sqrt{2}\sqrt{2+\Kb \left(\SpaceFunc-1\right)}-2-\Kb\left(\SpaceFunc-1\right)},\label{eqn:rd-of-R-one}\\
	r'(\SpaceFunc) =& \frac{-\Kb r(\SpaceFunc)/2\SpaceFunc}{2+\Kb\left(\SpaceFunc-1\right)+\sqrt{2}\sqrt{2+\Kb \left(\SpaceFunc-1\right)}}.\label{eqn:rd-of-R-two}
\end{align}
\end{subequations}
Substituting~\cref{eqn:branch-relation} into~\cref{eqn:rd-of-R-two}, we find that it transforms back into~\cref{eqn:rd-of-R-one}, confirming that the proposed relation between the branches is consistent. 

The solution of the differential~\cref{eqn:rd-of-R-one} corresponds to the vacuum solution being used in the main text, whilst the solution of~\cref{eqn:rd-of-R-two} is a new branch, with different properties. We can understand more about these properties by extending the results to include the metric time component~$\TimeFunc$. An explicit expression for~$\TimeFunc$ can be derived from~\cref{eq:rreom} with~$\mathrm{d}\TimeFunc(r)/\mathrm{d}r=\left[\mathrm{d}\TimeFunc(\SpaceFunc)/\mathrm{d}\SpaceFunc\right]\times\left[1/r'(\SpaceFunc)\right]$, where~$r'(\SpaceFunc)$ can be replaced with~\cref{eqn:rd-of-R-one}. After integration, one can get
\begin{align}
	&\TimeFunc_{\rm normal}(\SpaceFunc) =\nonumber\\
	&\hspace{30pt}\left(\frac{\sqrt{2+\Kb \SpaceFunc-\Kb}-\sqrt{2-\Kb}}{\sqrt{2+\Kb \SpaceFunc-\Kb}+\sqrt{2-\Kb}}\right)^{-\sqrt{\frac{2}{2-\Kb}}}.\label{eq:time_norm}
\end{align}
On the other hand, with the alternative choice for~$r'$ which one is led to by~\cref{eqn:rd-of-R-two}, and which can be generated from the first choice by using~\cref{eqn:branch-relation}, one gets the interesting result
\begin{equation}
\TimeFunc_{\rm alternate}(\SpaceFunc)=\frac{1}{\TimeFunc_{\rm normal}(\SpaceFunc)}.
\end{equation}
Thus there is a form of reciprocity in action for both~$\TimeFunc$ and~$r$ here, when written in terms of~$\SpaceFunc$. 

Note that both~$\TimeFunc_{\rm normal}$ and~$\TimeFunc_{\rm alternate}$ are set up so that they have the value 1 at the throat (i.e.\ where~$\SpaceFunc \mapsto \infty$), but any constant multiple of these can be used instead. In particular one might think that since~$\TimeFunc_{\rm normal}$ tends to a constant at spatial infinity, and can therefore be renormalised to be 1 at infinity, in keeping with wanting flat space there, then the same will apply to~$\TimeFunc_{\rm alternate}$. However, the relationship between~$r$ and~$\SpaceFunc$ is different in this case, and although the throat is still at~$\SpaceFunc \mapsto \infty$, we get to spatial infinity (i.e.~$r=\infty$) by letting~$\SpaceFunc \mapsto 0$ rather than~$\SpaceFunc \mapsto 1$, which is how spatial infinity in the normal case is reached. This has the effect that~$\TimeFunc \mapsto 0~$ as~$r \mapsto \infty$ in the `alternate' case, meaning that we cannot normalise it to be 1 at infinity. The space surrounding an object in which for some reason the alternate~$r$ and~$\TimeFunc$ solutions are picked out, is therefore pathological.

\subsection{An exact analytic solution for matter}\label{app:unphysical}

We now look at a case which although interesting analytically, does indeed find itself matched to an external space of the type just described.

This is based on the ansatz~$\Kb = 1/2$ and~$1/\SpaceFunc = 1-4\pi \rho_0 r^2/3$, where~$\rho_0$ is a constant. We note particularly that although this looks like one of the `extremal Buchdahl' cases discussed in the main text, in fact the choice of coefficient for~$r^2$, together with the choice of~$1/2$ for~$\Kb$, leads to a special case which stands outside the relations which lead to the solutions discussed in~\cref{sec:const_m_r}. 

In this ansatz, we find that the density is constant, with the value~$\rho_0$, and the pressure is given by
\begin{align}
    P = -\frac{\rho_0\left(5-8\pi r^2\rho_0\right)}{2\left(3-4\pi r^2\rho_0\right)}.
    \label{eq:neg_pres}
\end{align}
This goes to 0 at the radius~$r_\text{b}=\sqrt{5/(8\pi\rho_0)}$, which gives the radius of the object, and hence~$P$ is able to satisfy the boundary matching condition which we established above. Furthermore, although there is a throat according to the~$\SpaceFunc$ value, at~$r=\sqrt{3/(4\pi\rho_0)}$, this lies outside the radius~$r_b$, and so is not a problem, given that by this radius we should be using an external vacuum solution. 

This matter solution is therefore quite interesting, as being extremely simple analytically. However, once we do attempt to match to an external vacuum, one finds that it matches to the `alternate' branch just discussed, and therefore does not behave properly as~$r \mapsto \infty$. A speculation is that this is due to the fact that the pressure inside, though obeying the null, weak and dominant energy conditions for a perfect fluid, does violate the strong energy condition, in going from~$-(5/6) \rho_0$ at the centre to 0 at the boundary, and so for which~$\rho+3P$ fails to be~$\geq 0$ over some of the range. It will be of interest to investigate this in more general scenarios.

In this context, we notice that the `extremal Buchdahl' cases discussed in~\cref{sec:const_m_r}, despite having some properties which render them unphysical, such as infinite negative density at the origin, accompanying the infinite positive pressure there, nevertheless satisfy the strong energy criterion. (This is because the absolute value of the density is less than that of the pressure, over the range where the density is negative.) In this connection it is of interest that they do sit inside an asymptotically flat space that looks like Schwarzschild at infinity. This is for~$\Kb<1/2$. For~$\Kb>1/2$, the radius where~$\SpaceFunc \mapsto \infty$ sits \textit{inside} the object, and these solutions no longer work, as found already in~\cref{sec:const_m_r}.

\end{document}